\documentclass[11pt,a4paper]{article}

\usepackage{jheppub}

\usepackage{epsfig}
\usepackage{amssymb,amsmath}

\title{PDF reweighting in the Hessian matrix approach}

\author[a,b]{Hannu Paukkunen}
\author[c]{and Pia Zurita}

\affiliation[a]{Department of Physics, University of Jyv\"askyl\"a, P.O. Box 35, FI-40014 University of Jyv\"askyl\"a, Finland}
\affiliation[b]{Helsinki Institute of Physics, P.O. Box 64, FI-00014 University of Helsinki, Finland}
\affiliation[c]{Departamento de F\'\i sica de Part\'\i culas and IGFAE, Universidade de Santiago de
Compostela, E-15782 Galicia, Spain}

\emailAdd{hannu.paukkunen@jyu.fi}
\emailAdd{pia.zurita@usc.es}

\abstract{
We introduce the Hessian reweighting of parton distribution functions (PDFs). Similarly to the
better-known Bayesian methods, its purpose is to address the compatibility of new data and the
quantitative modifications they induce within an existing set of PDFs. By construction, the
method discussed here applies to the PDF fits that carried out a Hessian error analysis using a
non-zero tolerance $\Delta\chi^2$. The principle is validated by considering
a simple, transparent example. We are also able to establish an agreement with the Bayesian 
technique provided that the tolerance criterion is appropriately accounted for and that a
purely exponential Bayesian likelihood is assumed.
As a practical example, we discuss the inclusive jet production at the LHC. 
}

\begin{document}

\maketitle

\section{Introduction}

A large part of the present high-energy collider physics depends, in one way or another, on the
 knowledge of the parton distribution functions (PDFs). The use of PDFs
leans on a cornerstone theorem of Quantum Chromo Dynamics (QCD), the collinear factorization \cite{Collins:1989gx,Collins:1985ue}.
Although it can be formally proven only for the most simple cases, it is often assumed to work
in general. Ultimately, it is the agreement with the experimental data that decides whether such
an assumption is valid. 

The PDFs are traditionally determined in global analyses \cite{Forte:2013wc,Lai:2010vv,Martin:2009iq,Ball:2012cx}
finding the parametrization that can optimally reproduce a variety of experimental data. This is a complex
procedure requiring the ability to efficiently solve the parton evolution equations and to calculate
higher-order QCD cross-sections. Nontrivial issues are also the actual way of finding the best fit
and quantifying its uncertainties. Although various PDF parametrizations are publicly available for a general user,
for a long period of time it was difficult for e.g. an experimental collaboration to understand what
would be the implications of their measurements in the context of a global PDF fit.
For example, although a given measurement would be known to be most sensitive to, say, the
up quark distribution, some other data in a global fit may already provide the up quarks with 
more stringent constraints, and the real advantage of the measurement would be due to a sub-leading 
contribution, say, from the strange quarks. 

The Bayesian reweighting technique, first introduced in \cite{Giele:1998gw} and later on elaborated by the
Neural Network PDF (NNPDF) collaboration \cite{Ball:2010gb,Ball:2011gg}, provides a way of addressing the
consistency and quantitative effects of a new experimental evidence in terms of PDF fits. In essence, the 
underlying probability distribution, represented in the NNPDF philosophy \cite{Ball:2012cx} by an ensemble
($\sim 1000$) of PDF replicas, is updated by assigning each replica a certain weight based on the new data.
This method has become an increasingly popular way to estimate the effects of e.g. new LHC measurements
\cite{Chatrchyan:2013mza,Gauld:2013aja,Czakon:2013tha,Carminati:2012mm,Beneke:2012wb,d'Enterria:2012yj,Ball:2013hta}.
The drawback is that it has been proven to work only in conjunction with the NNPDF fits while
the majority of the existing PDF fits use a rather different way of quantifying the PDFs and their uncertainties.
Along with the best fit found by $\chi^2$ minimization, they provide a collection ($\sim 50$)
of Hessian error sets \cite{Pumplin:2001ct} that quantify the neighborhood of the central fit within a certain confidence
criterion $\Delta \chi^2$. An extension of the Bayesian reweighting technique to this particular case was
suggested in \cite{Watt:2012tq}, and has thereafter been used in some occasions \cite{Gauld:2013aja,Sato:2013wea,Armesto:2013kqa}.
However, a recent study \cite{Watt:2013oha} revealed clear deviations when comparing the results from reweighting
to the ones obtained by a direct fit, indicating that the proposed generalization is not accurate.

Here we take a different strategy. Based on the ideas presented in Ref.~\cite{Paukkunen:2013grz}, our principal
goal is to show how a general user can directly study the consistency and consequences of a new data
set within an existing set of PDFs that comes with the Hessian error sets without having to lean on
the Bayesian techniques. The method naturally incorporates the confidence criterion $\Delta \chi^2$ defined 
in the original fit and, by considering a simple numerical example, we argue that it is perfectly compatible
with a new fit. Our second objective is to understand how the procedure suggested here relates to the Bayesian reweighting
and to the discrepancies found in \cite{Watt:2013oha}.
We prove that the original Bayesian method, proposed in \cite{Giele:1998gw} and advocated recently in \cite{Sato:2013ika},
is equivalent with the one introduced here once the $\Delta \chi^2$ criterion is properly incorporated.

\section{The Hessian method}

The usual definition of an optimum correspondence between data and a set of PDFs $f\equiv f(x,Q^2)$
that depends on certain fit parameters $\{a\}$, is the minimum of a $\chi^2$-function. In its most simple form, we can write it as
\begin{equation}
 \chi^2\{a\} = \sum_k \left[ \frac{X_k^{\rm theory}[f] - X_k^{\rm data}}{\delta_k^{\rm data}} \right]^2,
\end{equation}
where $X_k^{\rm theory}[f]$ are the theory predictions depending on the PDFs. The corresponding
experimental values are denoted by $X_k^{\rm data}$ and their uncertainty by $\delta_k^{\rm data}$.
Modifications to this definition are necessary if the experimental errors are correlated or
some data sets are emphasized in the fit by assigning them an additional weight. 
In the Hessian 
approach to quantify the PDF errors \cite{Pumplin:2001ct}, the behaviour of $\chi^2$ around the 
best fit $S_0$  is approximated by a second order polynomial in the space of fit parameters $\{a\}$
\begin{equation}
 \chi^2\{a\} \approx \chi^2_0 + \sum_{ij} \delta a_i H_{ij} \delta a_j, \label{eq:chi2orig}
\end{equation}
where $\delta a_j \equiv a_j-a_j^0$ are the excursions from the best-fit values
and $\chi^2_0$ is the minimum value of $\chi^2$. Being symmetric, the Hessian 
matrix $H_{ij}$ has ${N_{\rm eig}}$ orthonormal eigenvectors ${\bf v}^{(k)}$ and  eigenvalues $\epsilon_k$
satisfying
\begin{eqnarray}
H_{ij} v_j^{(k)} & = & \epsilon_k v_i^{(k)} \, , \\
\sum_j v_j^{(k)} v_j^{(\ell)} & = & \sum_j v_k^{(j)} v_\ell^{(j)} = \delta_{k\ell}.
\end{eqnarray}
Defining a new set of variables as
\begin{equation}
 z_k \equiv \sqrt{\epsilon_k} \sum_j v_j^{(k)} \delta a_j, \label{eq:diag}
\end{equation}
one easily finds that
\begin{equation}
 \chi^2\{a\} \approx \chi^2_0  + \sum_i  z_i^2. \label{eq:chi2diagonalized}
\end{equation}
That is, the transformation in Eq.~(\ref{eq:diag}) diagonalizes the Hessian matrix.
A criterion is needed to specify 
how much the term $\sum_i  z_i^2$ can grow while the corresponding PDFs still remain ``acceptable''.
Those PDF fits that employ the ideal choice $\Delta \chi^2=1$ are usually limited a smaller set of
data \cite{CooperSarkar:2011aa,Alekhin:2012ig}, while the global fits prefer to take $\Delta \chi^2>1$
\cite{Lai:2010vv,Martin:2009iq} 
to account for small inconsistencies among the data sets and to compensate for the parametrization bias \cite{Pumplin:2009bb}.
It follows \cite{Pumplin:2001ct} that the corresponding uncertainty for a PDF-dependent 
quantity $\mathcal{O}=\mathcal{O}[f]$ can be computed as
\begin{eqnarray}
(\Delta \mathcal{O})^2 & = & {\Delta \chi^2} \sum_k \left( \frac{\partial \mathcal{O}}{\partial z_k} \, \right)^2 \, . \label{eq:err}
\end{eqnarray}
An essential feature of the Hessian approach is the introduction of the PDF error sets $S_k^\pm$,
defined customarily (along with the best fit $S_0$) in the $z$-space as
\begin{eqnarray}
z({S_0}) & = & \left(0,0,...,0 \right), \nonumber \\
z({S^\pm_1}) & = & \pm \sqrt{\Delta \chi^2} \left(1,0,...,0 \right), \label{eq:errset} \\
z({S^\pm_2}) & = &  \pm \sqrt{\Delta \chi^2} \left(0,1,...,0 \right), \nonumber \\
         & \vdots & \nonumber \\
z({S^\pm_{N_{\rm eig}}}) & = &  \pm \sqrt{\Delta \chi^2} \left(0,0,...,1 \right) \nonumber \, .
\end{eqnarray}
Using these sets, one can evaluate the derivatives in Eq.~(\ref{eq:err}) by a linear approximation 
\begin{equation}
\left( \frac{\partial \mathcal{O}}{\partial z_k}\right) \approx \frac{\mathcal{O}\left[{S_k^+} \right]
- \mathcal{O}\left[{S_k^-} \right]}{2\sqrt{\Delta \chi^2}},
\end{equation}
such that
\begin{eqnarray}
(\Delta \mathcal{O})^2 = \frac{1}{4} \sum_k \left( \mathcal{O}\left[{S_k^+} \right] - \mathcal{O}\left[{S_k^-} \right] \right)^2. \label{eq:mastererror}
\end{eqnarray}
This formula, or a generalization for asymmetric errors, provides an extremely simple and 
useful recipe for propagating the PDF-uncertainties to observables. Recently, it has become
fashionable \cite{Lai:2010vv,Martin:2009iq} to abandon the fixed $\Delta\chi^2$ tolerance
and define the PDF error sets instead by a ``dynamic tolerance''
\begin{equation}
 z_i(S_k^\pm) \equiv \pm t^\pm_k \delta_{ik},
\end{equation}
which coincides with Eq.~(\ref{eq:errset}) if the tolerance parameters $t^\pm_k$ are equal to $\sqrt{\Delta\chi^2}$.

\section{The Hessian reweighting}

Let us now consider a new set of data $\vec y=y_1,y_2,\ldots,y_{N_{\rm data}}$ with covariance matrix $C$.
Our goal here is twofold: to find out whether these new data are consistent within the original set of PDFs
and, if so, what would be the effect of incorporating them into the original analysis. In order to
answer these questions, we consider a function $\chi^2_{\rm new}$ defined as
\begin{equation}
 \chi^2_{\rm new} \equiv \chi^2_0  +   \sum_k^{N_{\rm eig}} z_k^2  + 
 \sum_{i,j=1}^{N_{\rm data}} \left(y_i[f]-y_i\right) C_{ij}^{-1} \left(y_j[f]-y_j\right), \label{eq:newchi2}
\end{equation}
where we have simply added the contribution of the new data on top of the ``old'' $\chi^2$ in Eq.~(\ref{eq:chi2diagonalized}). 
Using a similar linear approximation as earlier, we can estimate the theoretical values $y_i \left[f \right]$ 
in arbitrary $z$-space coordinates by
\begin{equation}
 y_i \left[f \right] \approx y_i \left[{S_0} \right] + \sum_{k=1}^{N_{\rm eig}} \frac{\partial y_i \left[{S} \right]}{\partial z_k}{\Big|_{S=S_0}} z_k
                   \approx y_i \left[S_0 \right] + \sum_{k=1}^{N_{\rm eig}} D_{ik} w_k, \label{eq:XS}
\end{equation}
where we have defined
\begin{eqnarray}
D_{ik} & \equiv & \frac{y_i\left[S_k^+ \right] - y_i\left[S_k^- \right]}{2} \label{eq:D} \\
w_k   & \equiv & \frac{z_k}{\frac{1}{2}\left(t_k^+ + t_k^-\right)}. \label{eq:w}
\end{eqnarray}
Thus, $\chi^2_{\rm new}$ is a continuous, quadratic function of the parameters $w_k$, and its
minimum is given simply by
\begin{equation}
{\vec {\bf w}^{\rm min}} = -{\bf B}^{-1} \vec {\bf a}, \label{eq:wmineq} 
\end{equation}
where the matrix ${\bf B}$ and vector $\vec {\bf a}$ are 
\begin{eqnarray}
 B_{kn} & = & \sum_{i,j} {D_{ik} C_{ij}^{-1} D_{jn}} + \left(\frac{t_k^+ + t_k^-}{2}\right)^2 \delta_{kn} \, , \label{eq:B} \\
 a_k    & = & \sum_{i,j} {D_{ik} C_{ij}^{-1} \left( y_j\left[S_0\right] - y_j \right)}. \label{eq:a} 
\end{eqnarray}
An important feature of the solution is the ``penalty term''
\begin{equation}
P \equiv \sum_{k=1}^{N_{\rm eig}} \left[\left(\frac{t_k^+ + t_k^-}{2}\right) w_k^{\rm min} \right]^2  \label{eq:pena}
\xrightarrow{t_k^\pm \rightarrow \sqrt{\Delta \chi^2}} \Delta \chi^2 \sum_{k=1}^{N_{\rm eig}} (w_k^{\rm min})^2,
\end{equation}
which we can use to decide whether the new data set is consistent within the original PDFs.
If an overall tolerance $\Delta \chi^2$ was defined in the original fit, having $P \ll \Delta \chi^2$ means
that the new data could have been incorporated into the original fit without causing a conflict with the other data.
On the other hand, if $P \gtrsim \Delta \chi^2$ the new data appears to display significant tension with the considered
set of PDFs. In the case that the original fit used a dynamic tolerance a simple interpretation like this is lost,
and one can only check whether the new $z$-space coordinates remain within the tolerance parameters.
However, even a large penalty term does not necessarily mean that the new data would be incompatible
with the other data. A situation like this may arise if the new data probe unconstrained components
of PDFs whose behaviour was fixed by hand. For example, some recent PDF fits \cite{Owens:2012bv} still 
assume $s(x) \propto (u(x)+d(x))$ for the strange quark distribution and confronted
with data sensitive to the strange quarks could lead to this kind of situation. 

The components of the weight vector $\vec {\bf w}^{\rm min}$ also specify the set
of PDFs $f^{\rm new}$ that corresponds to the new global minimum. They can be easily calculated by
taking $y_i=f(x,Q^2)$ in Eq.~(\ref{eq:XS}). That is,
\begin{equation}
 f^{\rm new} \approx f_{S_0} + \sum_{k=1}^{N_{\rm eig}} \left( \frac{f_{S^+_k}-f_{S^-_k}}{2} \right) w^{\rm min}_k. \label{eq:newPDF}
\end{equation}
The resulting new PDFs are linear combinations of the original ones --- they have been ``reweighted''. We note that
the new PDFs constructed in this way still satisfy the necessary sum rules. For
instance, as the original best fit $S_0$ and the error sets $S_k^\pm$ satisfy the
momentum sum rule
\begin{equation}
 \int_0^1 dx x \sum_f f_{S_0} = \int_0^1 dx x \sum_f f_{S_k^\pm} = 1,
\end{equation}
then 
\begin{eqnarray}
 \int_0^1 dx x\sum_f f^{\rm new} & = & \int_0^1 dx x \sum_f f_{S_0} 
 +  \sum_k \frac{w_k^{\rm min}}{2} \left[ \int_0^1 dx x \sum_f f_{S_k^+} - \int_0^1 dx x \sum_f f_{S_k^-} \right] \nonumber \\
& = & 1  + \sum_k \frac{w_k^{\rm min}}{2} \left[ 1 -1 \right] = 1. \nonumber
\end{eqnarray}
Due to the linearity of the parton evolution equations, also $f^{\rm new}$ satisfies them. Thus, the reweighted
distributions comprise  a proper set of PDFs which can be consistently utilized in perturbative QCD calculations.
One can also construct the new PDF error sets. Indeed, Eq.~(\ref{eq:newchi2}) can be rewritten as
\begin{equation}
 \chi^2_{\rm new} = {\chi^2_{{\rm new}}}_{\big|_{\vec{\bf w}=\vec{\bf w}^{\rm min}}}  +  \sum_{ij} \delta w_i B_{ij} \delta w_j, \label{eq:newchi2min}
\end{equation}
where ${\vec {\bf \delta w}} = {\vec {\bf w}} - {\vec {\bf w}_{\rm min}}$, and the matrix ${\bf B}$ takes the role
of the Hessian matrix (compare to Eq.~(\ref{eq:chi2orig})). This can be brought into a diagonal form by an
analogue of the transformation in Eq.~(\ref{eq:diag}),
and the new error sets defined as
\begin{equation}
\delta w_i(\hat S_k^\pm) = \pm \hat v_i^{(k)} \sqrt{\frac{1}{\hat\epsilon_k}} \hat t_k^\pm,
\end{equation}
where $\hat v_i^{(k)}$ are the eigenvectors and $\hat \epsilon_k$ the eigenvalues of the matrix $\bf B$.
The original overall tolerance can be set easily by $\hat t_k^\pm=\sqrt{\Delta \chi^2}$.

\subsection*{Non-linear extension of the Hessian reweighting} 
\label{sec:nonlin}

The linear approximation of Eq.~(\ref{eq:XS}) can be improved by including also quadratic
terms of $w_k$ as
\begin{eqnarray}
y_i\left[S \right] = y_i\left[S_0 \right] & + & \sum_{k=1}^{N_{\rm eig}} \frac{1}{2}
\left[ 
\frac{y_i\left[S_k^+ \right]- y_i\left[S_0 \right]}{t^+_k/t^-_k}  -
\frac{y_i\left[S_k^- \right]- y_i\left[S_0 \right]}{t^-_k/t^+_k} 
 \right]w_k
\label{eq:nonliny} \\
& + & \sum_{k=1}^{N_{\rm eig}} \frac{t^+_k+t^-_k}{4}
\left[ 
\frac{y_i\left[S_k^+ \right]- y_i\left[S_0 \right]}{t^+_k} 
 +
\frac{y_i\left[S_k^- \right]- y_i\left[S_0 \right]}{t^-_k} 
 \right]w_k^2 \nonumber
\end{eqnarray}
correcting for the possible non-linear behaviour.
Using this in Eq.~(\ref{eq:newchi2}), $\chi^2_{\rm new}$ becomes a quartic function of $w_k$
whose
minimum must be found by numerical methods. The corresponding PDFs can be computed by taking $y_i=f(x,Q^2)$ in
Eq.~(\ref{eq:nonliny}). The matrix ${\bf B}$ in Eq.~(\ref{eq:newchi2min}) gets replaced by
\begin{eqnarray}
 B^{\rm non.lin}_{kn} & = & \sum_{i,j} \left( \frac{\partial y_i[f]}{\partial w_k} \right) C^{-1}_{ij} \left( \frac{\partial y_j[f]}{\partial w_n} \right) 
 \\  & + & \sum_{i,j} \left( \frac{\partial^2 y_i[f]}{\partial w_k \partial w_n} \right) C^{-1}_{ij} \left(y_j[f]-y_j\right)
+ \left(\frac{t_k^+ + t_k^-}{2}\right)^2 \delta_{kn}, \nonumber          
\end{eqnarray}
where the partial derivatives read
\begin{eqnarray}
\left( \frac{\partial y_i[f]}{\partial w_k} \right) & = & 
\frac{1}{2} \left[ 
\frac{y_i\left[S_k^+ \right]- y_i\left[S_0 \right]}{t^+_k/t^-_k}  -
\frac{y_i\left[S_k^- \right]- y_i\left[S_0 \right]}{t^-_k/t^+_k} 
 \right] \\
& + &  \frac{t^+_k+t^-_k}{2}
\left[ \frac{y_i\left[S_k^+ \right]- y_i\left[S_0 \right]}{t^+_k} 
 + \frac{y_i\left[S_k^- \right]- y_i\left[S_0 \right]}{t^-_k} \right]w_k \nonumber
 \\
\left( \frac{\partial^2 y_i[f]}{\partial w_k \partial w_n} \right) & = &
\frac{t^+_k+t^-_k}{2}
\left[ \frac{y_i\left[S_k^+ \right]- y_i\left[S_0 \right]}{t^+_k} 
 + \frac{y_i\left[S_k^- \right]- y_i\left[S_0 \right]}{t^-_k} \right]
 \delta_{kn},
\end{eqnarray}
and are understood to be evaluated at the found minimum.

\section{Bayesian methods}
\label{sec:Bayes}

Given a large ensemble of PDFs $f_k$, $k=1 \ldots N_{\rm rep}$, such as those of the  NNPDF
collaboration~\cite{Ball:2012cx}, that represents the underlying probability
distribution $\mathcal{P}_{\rm old}(f)$ of the PDFs, one can compute the expectation value
$\langle \mathcal{O} \rangle$ and variance $\delta \langle \mathcal{O} \rangle$ for an 
observable $\mathcal{O}$ as
\begin{eqnarray}
\langle \mathcal{O} \rangle  & = & \frac{1}{N_{\rm rep}} \sum_{k=1}^{N_{\rm rep}} \mathcal{O} \left[ f_k \right], \\
\delta \langle \mathcal{O} \rangle  & = & \sqrt{\frac{1}{N_{\rm rep}} \sum_{k=1}^{N_{\rm rep}} \left( \mathcal{O} \left[ f_k \right] - \langle \mathcal{O} \rangle\right)^2 }.
\end{eqnarray}
Using the laws of statistics, the initial probability distribution $\mathcal{P}_{\rm old}(f)$ 
can be updated to include also additional information contained in a new set of data $\vec y$,
since, by the Bayes theorem,
\begin{equation}
\mathcal{P}_{\rm new}(f) \varpropto \mathcal{P}(\vec y \vert f) \, \mathcal{P}_{\rm old}(f)\, ,
\label{eq:proba}
\end{equation}
where $\mathcal{P}(\vec y  \vert f)$ stands for the conditional probability (the likelihood function) 
for the new data, given a set of PDFs.
It follows that the average value for any observable depending on the PDFs becomes a weighted average
\begin{eqnarray}
\langle \mathcal{O} \rangle_{\rm new}  & = & \frac{1}{N_{\rm rep}} \sum_{k=1}^{N_{\rm rep}} \omega_k \, \mathcal{O} \left[ f_k \right], \label{eq:Bnew} \\
\delta \langle \mathcal{O} \rangle_{\rm new}  & = & \sqrt{\frac{1}{N_{\rm rep}} \sum_{k=1}^{N_{\rm rep}} \omega_k \, \left( \mathcal{O} \left[ f_k \right] - \langle \mathcal{O} \rangle_{\rm new} \right)^2 } \, , \label{eq:Bnew2}
\end{eqnarray}
where the weights $\omega_{k}$ turn out to be proportional to the likelihood function $\mathcal{P}(\vec y \vert f)$.
The question of how to choose the likelihood appropriately has been recently revisited in \cite{Sato:2013ika} but
the conclusive answer, if it exists, remains to be given. Two options, corresponding to different choices of the likelihood,
have been discussed in the literature. The one suggested originally by Giele and Keller (GK) \cite{Giele:1998gw} follows from
taking $\mathcal{P}(\vec y \vert f) d^ny$ as the probability to find the new data to be confined in a differential element
$d^n{y}$ around $\vec{y}$ resulting in
\begin{equation}
 \omega_k^{\rm GK} = \frac{\exp\left[-\chi^2_k/2 \right]}{(1/N_{\rm rep}) \sum_{k=1}^{N_{\rm rep}} 
\exp\left[-\chi^2_k/2 \right]}, \label{eq:wGK}
\end{equation}
where 
\begin{equation}
 \chi^2_k = \sum_{i,j=1}^{N_{\rm data}} \left(y_i[f_k]-y_i\right) C_{ij}^{-1} \left(y_j[f_k]-y_j\right). \label{eq:chi2onlynew}
\end{equation}
The option advocated by the NNPDF collaboration derives from taking $\mathcal{P}(\vec y \vert f) d\chi$
as the probability for the corresponding $\chi\equiv\sqrt{\chi^2}$ to be confined in a differential volume $d\chi$ around $\chi$, giving
instead\footnote{We dub these weights $\omega_k^{\rm chi-squared}$ since their behaviour 
is very close to the usual $\chi^2$ distribution.}
\begin{equation}
 \omega_k^{{\rm chi-squared}} = \frac{\left( \chi^2_k \right)^{(N_{\rm data}-1)/2} \exp\left[-\chi^2_k/2 \right]}{(1/N_{\rm rep}) \sum_{k=1}^{N_{\rm rep}} \left( \chi^2_k \right)^{(N_{\rm data}-1)/2} 
\exp\left[-\chi^2_k/2 \right]}, \label{eq:wNNPDF}
\end{equation}
which has been shown to be consistent with a direct fit in the NNPDF framework \cite{Ball:2010gb,Ball:2011gg}.
It was pointed out in Ref.~\cite{Sato:2013ika} that the former weights contain more information on the new data
than the latter ones, as a given data set uniquely determines the value of $\chi^{2}$, while a fixed
$\chi^{2}$ may correspond to various different data sets.
The generic behaviour of these weights with respect to $\chi^{2}$ per number of points for $N_{\rm data}=10$ is shown
\begin{figure}[ht!]
\begin{center}
\vspace*{-0.4cm}
\epsfig{figure=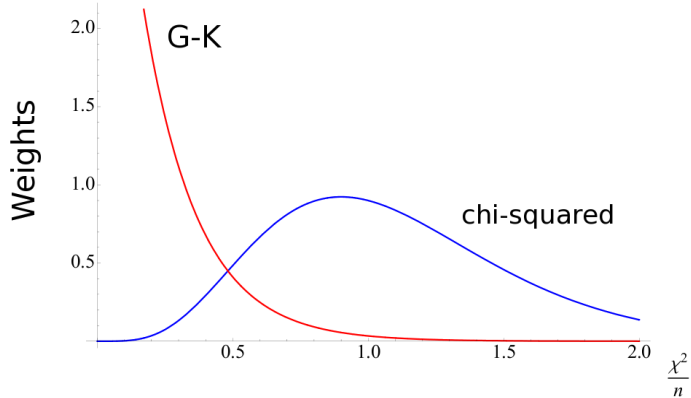,width=0.60\textwidth}
\end{center}
\vspace*{-0.7cm}
\caption{\label{fig:weights}
Comparison of the likelihoods for the original derivation of the Bayesian re-weighting \cite{Giele:1998gw} (red) 
and the one proposed in \cite {Ball:2010gb,Ball:2011gg} (blue). In this plot the number of points is $n=10$.}
\end{figure}
in Figure~\ref{fig:weights}. While the GK weights are always higher for those replicas that give lower $\chi^{2}$,
the NNPDF option obviously favors ones with $\chi^2/N_{\rm data} \approx 1$. We note that the latter likelihood
may lead to a following situation: If the value of $\chi^2/N_{\rm data}$ (computed with the expectation
values of the observables or as an average of the individual $\chi^2_k$s) is less than unity before the reweighting,
it can happen that the reweighting actually causes $\chi^2$ for the new data to grow since the replicas with 
$\chi^{2}/N_{\rm data} \approx 1$ are favored. However, if new data are directly included into a PDF fit
as in Eq.~(\ref{eq:newchi2}), the value of $\chi^2$ for the new data should only decrease.\footnote{
In PDF fits with extremely flexible fit functions, there is a danger that the PDFs that correspond to
the minimum $\chi^2$ unwantedly reproduce also random fluctuations of the data. Here, we assume that
this is not the case or that it has been taken into account e.g. by including suitable penalty terms 
in the original definition of $\chi^2$ \cite{Pumplin:2009bb,Glazov:2010bw}.
}

The large ensemble of PDFs required by the Bayesian approach can be constructed, in analogue
to Eq.~(\ref{eq:newPDF}), by
\begin{equation}
 f_k \equiv f_{S_0} + \sum_i^{N_{\rm eig}} \left( \frac{f_{S^+_i}-f_{S^-_i}}{2} \right) R_{ik} \label{eq:replicas},
\end{equation}
where the coefficients $R_{ik}$ are random numbers drawn from a Gaussian distribution centered at zero 
and with variance one.
An asymmetric version of Eq.~(\ref{eq:replicas}) to account for non-linearities was advocated
in Ref.~\cite{Watt:2012tq}. Specifically, it was proposed that the replicas should be generated by
\begin{equation}
f_k^{\rm asym} \equiv f_{S_0} + \sum_i^{N_{\rm eig}} \left( {f_{S^\pm_i}-f_{S_0}} \right) |R_{ik}| 
\end{equation}
where $f_{S^+_i}$ or $f_{S^-_i}$ is chosen according to the sign of $R_{ik}$. However, in this case already before
the reweighting the expectation values for the observables will not, in general, match those computed directly with the 
central set of the original fit. To accurately compare with the linear Hessian reweighting, we stick here to the symmetric
prescription of Eq.~(\ref{eq:replicas}).
As pointed out earlier, the replicas built in this way satisfy the
PDF sum rules and the parton evolution equations. After computing the weights $\omega_k$ for each replica,
the reweighted PDFs can be written as
\begin{equation}
 f_{\rm new} = f_{S_0} + \sum_i^{N_{\rm eig}} \left( \frac{f_{S^+_i}-f_{S^-_i}}{2} \right)
 \left ( \frac{1}{N_{\rm rep}} \sum_k^{N_{\rm rep}} \omega_k R_{ik} \right) \label{eq:MCres},
\end{equation}
and, similarly to the Hessian case, one can calculate the ``penalty'' induced in the original fit by
\begin{eqnarray}
P = \Delta \chi^2 \, \sum_i^{N_{\rm eig}} \left( \frac{1}{N_{\rm rep}} \sum_k^{N_{\rm rep}} \omega_k R_{ik} \right)^2. \label{eq:penaltyB}
\end{eqnarray}
We note that before reweighting $w_k=1$, and the sums in the parenthesis above vanish since the mean
of the random numbers $R_{ik}$ is zero.  Another useful indicator for the Bayesian methods is
the effective number of replicas $N_{\rm eff}$, defined as
\begin{equation}
N_{\rm eff} \equiv \exp \left\{ \frac{1}{N_{\rm rep}} \sum_{k=1}^{N_{\rm rep}} \omega_k \log(N_{\rm rep}/\omega_k)\right\}.
\end{equation}
If a given replica $f_k$ ends up having a small weight $w_k \ll 1$, it has
a negligible effect in the new predictions computed by Eqs.~(\ref{eq:Bnew}) and (\ref{eq:Bnew2}). The value
of $N_{\rm eff}$ defined above serves as an estimate for such a ``loss'' of replicas. If $N_{\rm eff} \ll N_{\rm rep}$,
the method becomes inefficient and is a sign that the new data contains too much new information or
that it is incompatible with the previous data. Should this happen, also the penalty in Eq.~(\ref{eq:penaltyB}) is probably large.

\vspace{-0.2cm}
\section{Simple example}
\label{Sec:SimpleExample}

\vspace{-0.2cm}
In this section, we will compare the different reweighting methods by invoking a rather
simple, but illustrative example. We consider a function
\begin{equation}
 g(x) = a_0 x^{a_1} (1-x)^{a_2}  e^{x a_3}  (1 + x e^{a_4})^{a_5},
\end{equation}
which resembles a typical fit function used in PDF fits.\footnote{In fact, the functional
form and the parameter values we use here correspond to the gluons of CTEQ6 \cite{Pumplin:2002vw}.}
We proceed as follows:
\begin{itemize}
 \item Construct a set of pseudodata (data set 1) for $g(x)$. The value of each data point $y_k$
and its uncertainty $\delta y_k$ are computed by
$$
y_k = (1+\alpha r_k)y_k^0, \quad \delta y_k = \alpha y_k^0
$$
where $y_k^0=g(x)$ is evaluated with fixed parameters $a_0^0=30$, $a_1^0=0.5$, $a_2^0=2.4$, $a_3^0=4.3$,
$a_4=2.4$, and $a_5=-3$. The parameter $\alpha=0.05$ controls the size of the fluctuations generated
by the Gaussian random numbers $r_k$.
\item Perform a $\chi^2$ fit with four free parameters $a_0$, $a_1$, $a_2 $, $a_3$ to these data,
and construct the corresponding Hessian error sets using a certain $\Delta \chi^2$ criteria.
\item Construct a second set of pseudodata (data set 2) by the same procedure as in the case
of data set 1 (using the same parameters), and apply the above-introduced reweighting techniques on these data.
\item Perform a direct fit using both data sets and compare this ``complete'' result to the
predictions given by the reweighting methods.
\end{itemize}

\begin{figure*}[ht!]
\begin{minipage}[b]{1.00\linewidth}
\centering
\includegraphics[width=1.00\textwidth]{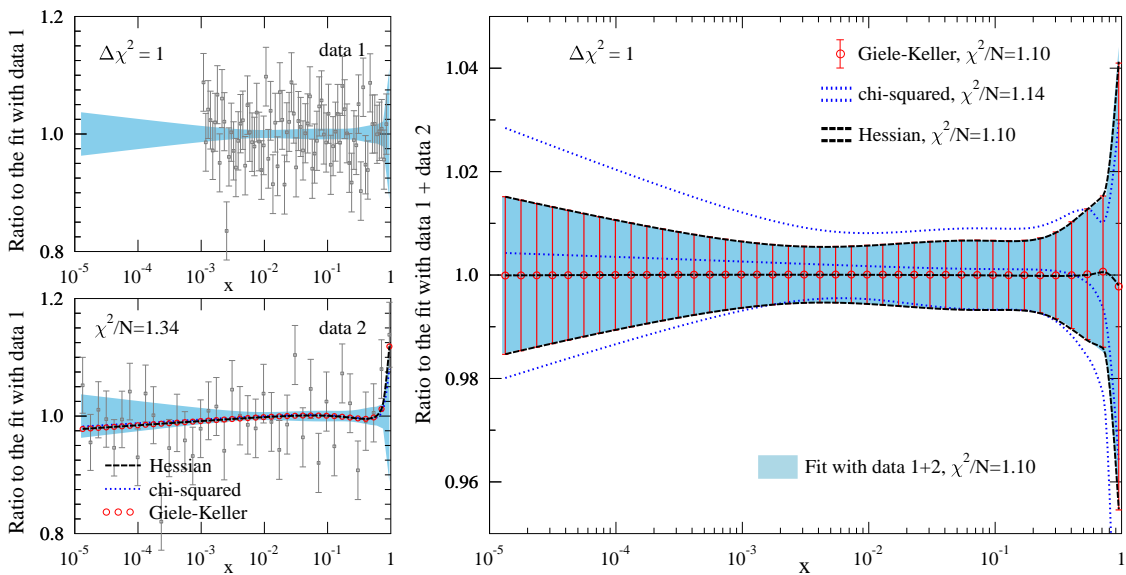}
\end{minipage}
\caption{{\bf Upper left-hand panel:} Data set 1 normalized by the fit to
these data. The light blue band shows the fit error defined by $\Delta \chi^2=1$.
{\bf Lower left-hand panel:} Data set 2 normalized by the fit to
the data set 1. The light blue band shows the original fit error and the black dashed
line is the result using the Hessian reweighting on these data. Blue
dotted line and the line marked by red circles are the corresponding results
using the Bayesian reweighting with $\omega_k^{{\rm chi-squared}}$ and $\omega_k^{{\rm GK}}$,
respectively. The original value of $\chi^2/N$ is indicated.
{\bf Right-hand panel:} The results of reweighting for the function $g(x)$ normalized to the fit using
data sets 1 and 2 (the light-blue band is the total $\Delta \chi^2=1$ error band).
The band enclosed by the black dashed lines corresponds the Hessian reweighting,
and the one enclosed by the blue dotted lines to the Bayesian reweighting with
$\omega_k^{{\rm chi-squared}}$. The red circles with error bars are the results using Bayesian
reweighting with $\omega_k^{{\rm GK}}$. The resulting values of $\chi^2/N$ for the data set 2
are indicated.
}
\label{Fig:example1}
\end{figure*}

\begin{figure*}[ht!]
\begin{minipage}[b]{1.00\linewidth}
\centering
\includegraphics[width=0.50\textwidth]{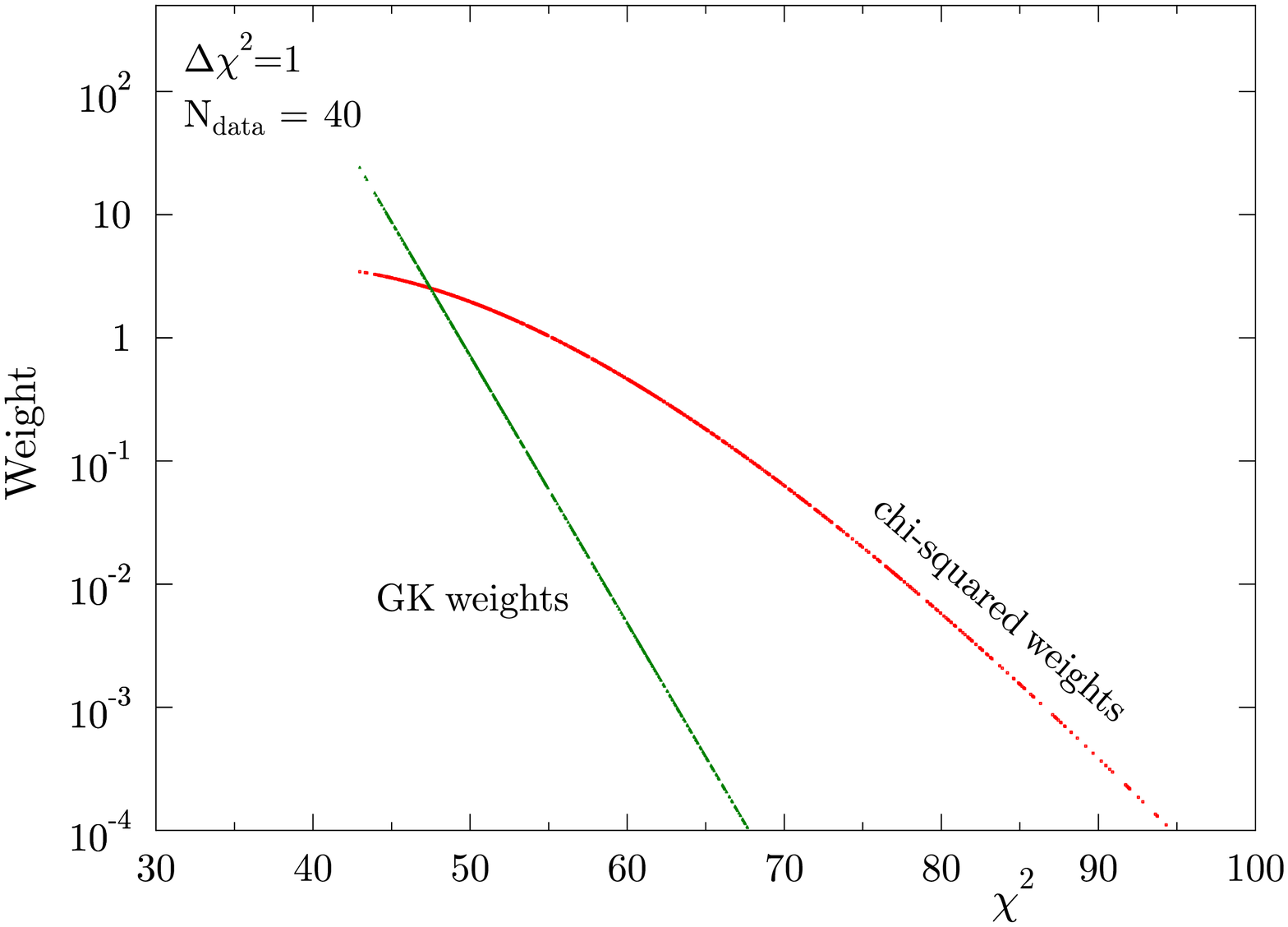}
\hspace{-0.6cm}
\includegraphics[width=0.50\textwidth]{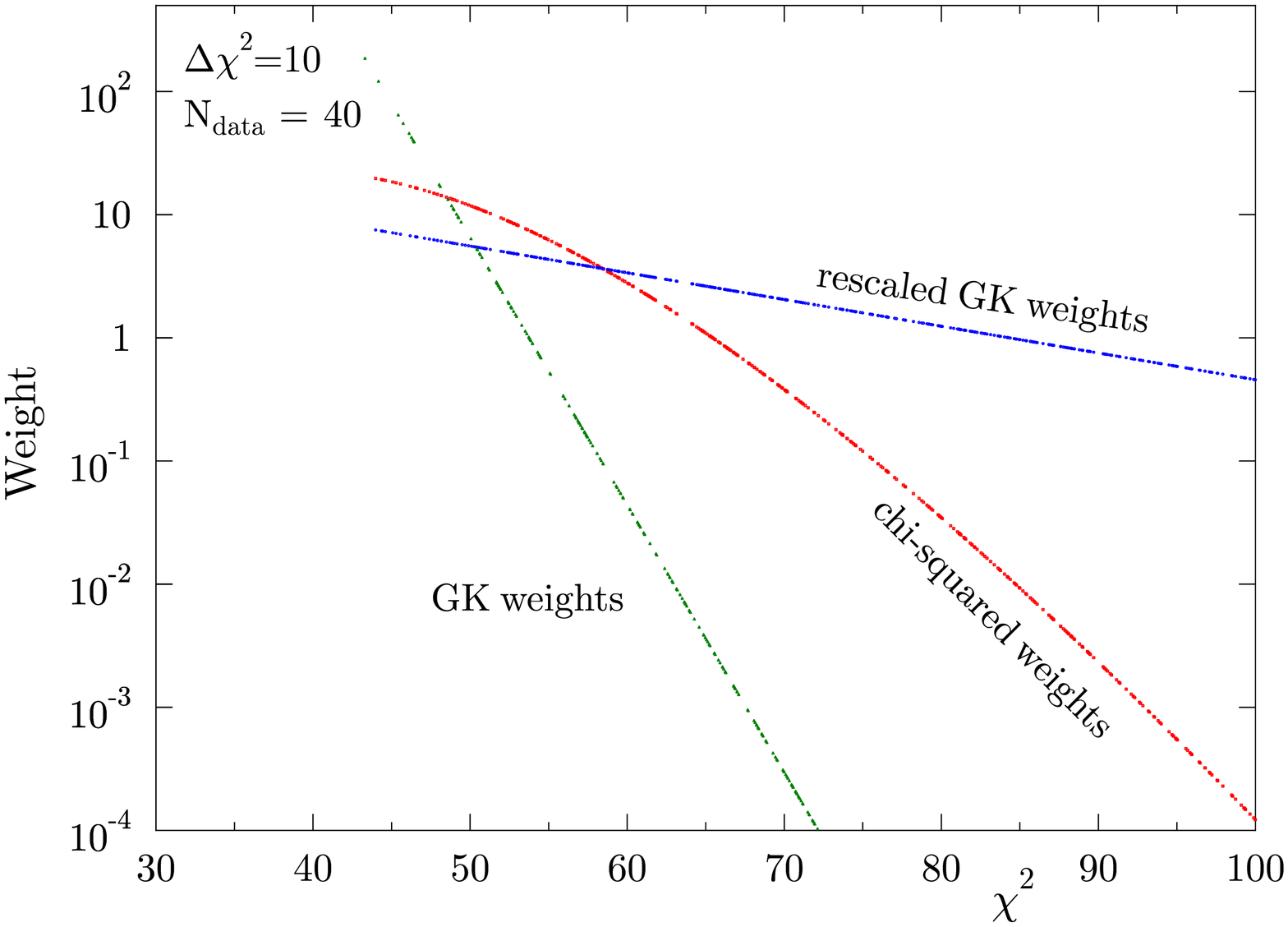}
\end{minipage}
\caption{{\bf Left-hand panel:} Distribution of the Bayesian weights for $\Delta \chi^2=1$ corresponding
to Figure~\ref{Fig:example1}. The line of green (red) points correspond to the GK (chi-squared) weights.
{\bf Right-hand panel:} As the left-hand panel but for $\Delta \chi^2=10$ corresponding to Figures~\ref{Fig:example2}
and \ref{Fig:example3}. The additional line of blue dots corresponds to the rescaled GK weights.}
\label{Fig:wdist}
\end{figure*}

We begin by considering the ideal case $\Delta \chi^2=1$ (that is, we take $t_k^\pm = 1$), depicted in Figure~\ref{Fig:example1}.
We have chosen here an example in which the data set 2 (40 points) contains evidence from a region of $x$
that the data set 1 (80 points) did not reach. In the case of Bayesian methods, we have used a
sufficiently large number ($10^5$) of replicas to get rid of all numerical inaccuracies. To compare
the methods as accurately as possible, the linear version of Hessian reweighting is used
throughout this section. The results shown in Figure~\ref{Fig:example1} reveal that the Hessian reweighting
and the Bayesian one with GK weights agree not only with each other, but also with the direct re-fit.
The outcome with the chi-squared weights is similar but it does not fully coincide with the others. The reason for the 
similarity is that as the pseudodata we have used here are statistically consistent and we do not have too much freedom
in the fit function, there are practically no replicas with $\chi_k^2/N_{\rm data} < 1$, 
as can be appreciated from the left-hand panel of Figure~\ref{Fig:wdist} where we plot the distribution
of both Bayesian weights for this particular case. That is,
the turnover for the chi-squared weights (see Figure~\ref{fig:weights}) does not play a role and
it is very similar subset of replicas that mostly contributes in both cases.

\begin{figure*}[t!]
\begin{minipage}[b]{1.00\linewidth}
\centering
\includegraphics[width=1.0\textwidth]{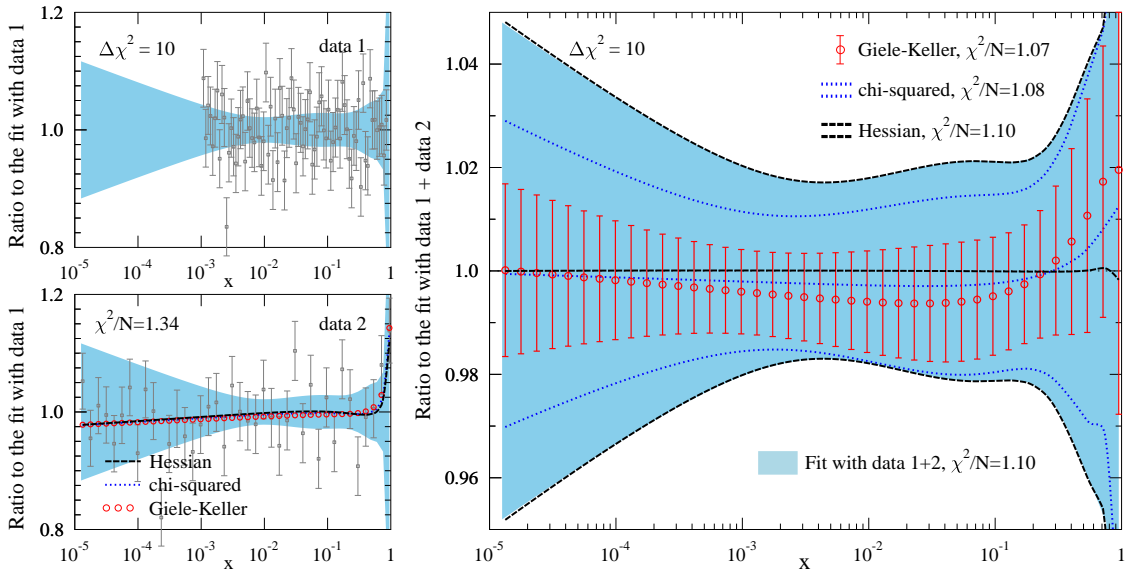}
\end{minipage}
\caption{As Figure~\ref{Fig:example1}, but using $\Delta \chi^2 = 10$.}
\label{Fig:example2}
\end{figure*}

Many global fits of PDFs use $\Delta \chi^2$ clearly larger than unity. In our simple example here,
a motivation for using $\Delta \chi^2 > 1$ could be to compensate for the restricted functional
form at small values of $x$ where the data set 1 did not have constraints \cite{Pumplin:2009bb}.
Thus, we repeat the exercise taking this time $\Delta \chi^2 = 10$. The results are shown in
Figure~\ref{Fig:example2}. While the Hessian reweighting can still accurately reproduce the
re-fit, neither of the Bayesian methods coincides with them. In the case of GK reweighting the reason for the
failure is that the likelihood function $\mathcal{P}(\vec y  \vert f)$ as such does not contain
any information on $\Delta \chi^2$, although the distribution of replicas clearly depends on the
value of $\Delta \chi^2$. 
As the spread among the replicas encoded by $\Delta \chi^2 = 1$ is narrower than that
covered by $\Delta \chi^2 = 10$, the new data appear more constraining than they actually are.
This can also be verified from the distribution of Bayesian weights shown in the right-hand panel
of Figure~\ref{Fig:wdist}. In comparison to the case with $\Delta \chi^2 = 1$ the replicas with lowest
$\chi^2$ tend to get much higer weight.
\begin{figure*}[ht!]
\begin{minipage}[b]{1.00\linewidth}
\centering
\includegraphics[width=1.00\textwidth]{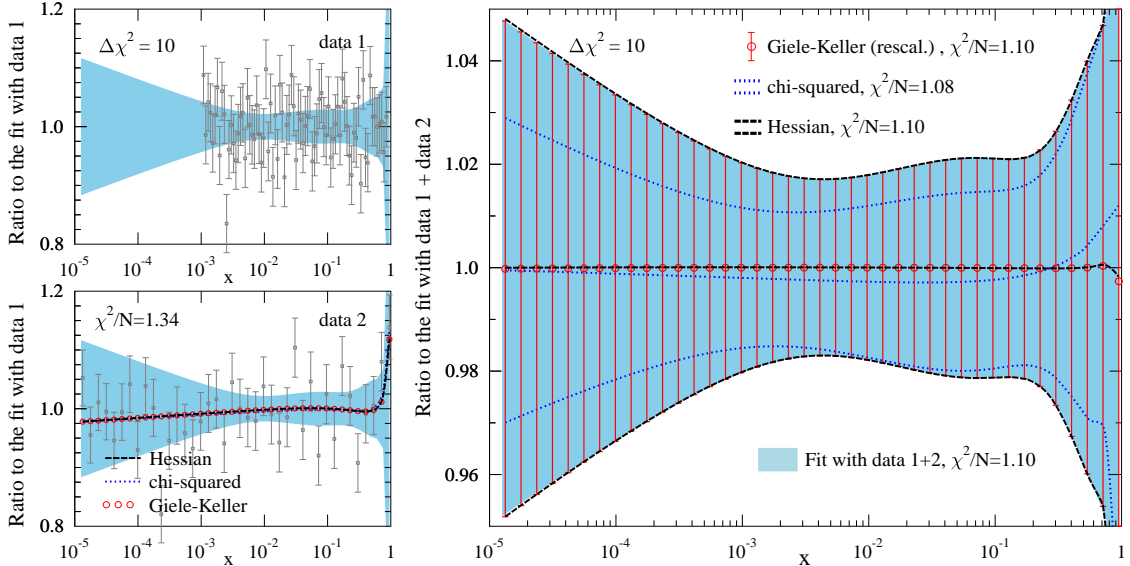}
\end{minipage}
\caption{As Figure~\ref{Fig:example2}, but rescaling the values of $\chi^2$ by $(\Delta \chi^2)^{-1}$
in the case of Bayesian reweighting with GK weights.}
\label{Fig:example3}
\end{figure*}
However, the agreement encountered with $\Delta \chi^2 = 1$ (Figure~\ref{Fig:example1}) hints
that it should be possible to generalize the Bayesian method with GK weights also to the case with $\Delta \chi^2 > 1$. The
key point is to note that we could divide, for example, Eq.~(\ref{eq:newchi2}) by $\Delta\chi^2$, and 
effectively use $\Delta\chi^2=1$ thereafter. This observation instructs us to rescale the values of 
$\chi^2_k$ in Eqs.~(\ref{eq:wGK}) as
\begin{equation}
 \chi^2_k \longrightarrow \frac{\chi^2_k}{\Delta \chi^2}, \label{eq:rescale}
\end{equation}
when computing the weight for each replica. The corresponding results are shown in
Figure~\ref{Fig:example3} which differs from Figure~\ref{Fig:example2} only in using 
the above mentioned rescaling when computing the Bayesian GK weights. Evidently, the 
agreement between the Bayesian method with GK weights, the Hessian reweighting, and 
the re-fit, is restored.
The spectrum of rescaled GK weights is shown in the right-hand panel of
Figure~\ref{Fig:wdist} as well. The division of the individual $\chi^2$ values
by $\Delta \chi^2$ makes the distribution of weights considerably flatter
and it is actually a rather wide range of $\chi^2$ values
that still give a non-negligible contribution. We note that mathematically we would
end up with the same reweighting result by narrowing the Gaussian distribution of random
numbers instead of rescaling the individual $\chi^2$ values. Indeed, replacing $R_{ik} \rightarrow R_{ik}/\sqrt{\Delta \chi^2}$
in Eq.~(\ref{eq:replicas}) we would recover the same distribution of replicas that was
obtained by using $\Delta \chi^2=1$. However, with $\Delta \chi^2=10$ and rescaled
GK weights the effective number of replicas is about twice as large as that with
$\Delta \chi^2=1$ ($N_{\rm eff,\Delta \chi^2=1} \approx 19200$,
$N_{\rm eff,\Delta \chi^2=10} \approx 37500$). This makes the rescaling of $\chi^2$ 
values a better option that narrowing the Gaussian distribution, although both procedures
lead to the same result.

Our simple example here indicates that the Bayesian reweighting with chi-squared weights does not, in general, 
correspond to a direct re-fit although here they lead to a good approximation.\footnote{In a 
situation in which many replicas give $\chi_k^2/N_{\rm data} \ll 1$ the difference to the direct re-fit can become much larger.} 
However, comparing Figures~\ref{Fig:example1} and \ref{Fig:example3} we notice that
for $\Delta \chi^2 = 1$ the results of chi-squared reweighting are a bit above 
the true result, but for $\Delta \chi^2 = 10$ somewhat below. Similarly, for $\Delta \chi^2 = 1$
the errorband is too wide, but for $\Delta \chi^2 = 10$ too narrow. It therefore looks
possible that by ``tuning'' the $\Delta \chi^2$ appropriately the chi-squared reweighting
could be made to coincide with the exact result. Since the rescaled GK weights appear to
be the proper way to do the reweighting, the condition that the replicas that mostly contribute
get the same weight in both cases  is that the ratio
\begin{equation}
r \equiv \frac{\left( \chi^2 \right)^{(N_{\rm data}-1)/2} \exp\left[-\chi^2/2 \right]}{\exp\left[-\chi^2/(2\Delta \chi^2) \right]} 
\end{equation}
is approximately constant. Requiring the derivative of this ratio to be zero, one finds
\begin{equation}
 \Delta \chi^2_{\big|{\frac{dr}{d\chi^2}=0}} = \frac{\chi^2}{\chi^2-(N_{\rm data}-1)}. \label{eq:opimumchi2}
\end{equation}
\begin{figure*}[ht!]
\begin{minipage}[b]{1.00\linewidth}
\centering
\includegraphics[width=1.00\textwidth]{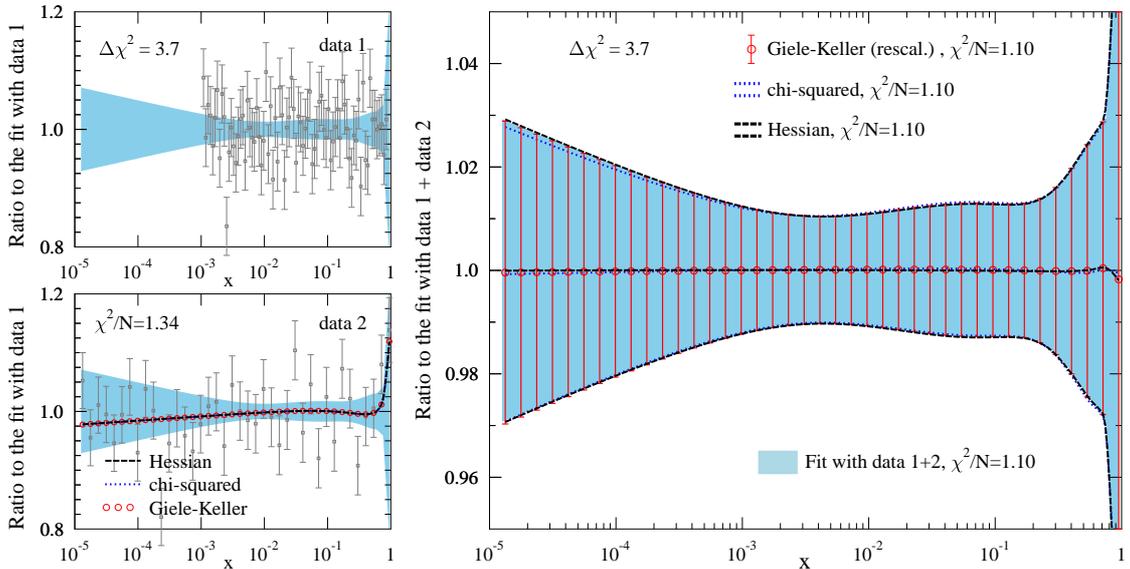}
\end{minipage}
\caption{As Figure~\ref{Fig:example3}, but using $\Delta \chi^2 = 3.7$.}
\label{Fig:example4}
\end{figure*}
\begin{figure*}[ht!]
\begin{minipage}[b]{1.00\linewidth}
\centering
\includegraphics[width=0.50\textwidth]{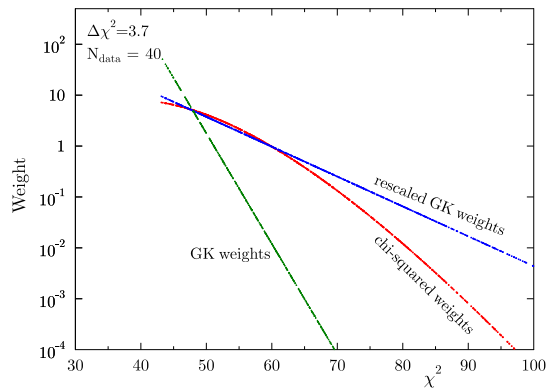}
\end{minipage}
\caption{As the right-hand panel of Figure~\ref{Fig:wdist}, but for $\Delta \chi^2=3.7$.}
\label{Fig:wdist2}
\end{figure*}
The individual replicas are always centered around the original fit and it thus makes sense to take
the original value of $\chi^2$ as a guideline in evaluating the value of $\Delta \chi^2$ in the equation above.
In the present example $\chi^2/N \approx 1.34$ before the re-fit which corresponds to $\Delta \chi^2 \approx 3.7$.
This particular value of $\Delta \chi^2$ brings the chi-squared method to an excellent agreement
with the direct re-fit as can be seen from Figure~\ref{Fig:example4}. The corresponding distributions
of the Bayesian weights are shown in Figure~\ref{Fig:wdist2} and, indeed, with $\Delta \chi^2 \approx 3.7$
the chi-squared and rescaled GK weights go practically hand in hand in the region of $\chi^2$ which mostly contributes
to the final result. That is, if most of the replicas give $\chi^2/N > 1$ and $\Delta \chi^2$ happens to be 
close to the value of Eq.~(\ref{eq:opimumchi2}) evaluated with the original central $\chi^2$, both Bayesian
reweightings may give approximately the same result.

In our simple example here, what mainly limits the accuracy of the reweighting
is the precision of the original quadratic expansion of Eq.~(\ref{eq:chi2orig}). 
Indeed, the small mismatch between the results of reweighting and the real fit e.g. in 
Figure~\ref{Fig:example3} can be largely attributed to this approximation not being perfect.
Although the same function $g(x)$ was used in generating and fitting the pseudodata, we have
checked that using a different fit function, e.g. a superposition of Chebysev polynomials,
would not alter our conclusions.

\section{Equivalence of the Bayesian and Hessian reweighting}

The close similarity of the results obtained using the (linear) Hessian reweighting and Bayesian one with the
rescaled GK weights indicates that the two are actually one and the same. In this short section we will give a formal
proof of this equivalence. From Eq.~(\ref{eq:MCres}) we see that the coordinates specifying the GK-reweighted
PDFs in the eigenvector space are given by
\begin{equation}
 w_k^{\rm GK} = \frac{1}{N_{\rm rep}} \sum_\ell^{N_{\rm rep}} \omega_\ell R_{k\ell}
              = \frac{1}{\mathcal{N}} \times \frac{1}{N_{\rm rep}} \sum_\ell^{N_{\rm rep}} e^{-\frac{\chi^2_\ell}{2\Delta \chi^2}}  R_{k\ell},
\end{equation}
where the denominator $\mathcal{N}$ is
\begin{equation}
 \mathcal{N} \equiv \frac{1}{N_{\rm rep}} \sum_\ell^{N_{\rm rep}} e^{-\frac{\chi^2_\ell}{2\Delta \chi^2}},
\end{equation}
and we have applied the rescaling $\chi^2_k \rightarrow {\chi^2_k}/{\Delta \chi^2}$ as in Eq.~(\ref{eq:rescale}).
Using the expression for $\chi^2$ in Eq.~(\ref{eq:chi2onlynew}) and the linear approximation
of Eq.~(\ref{eq:XS}) with $w_k = R_{k\ell}$, we find
\begin{eqnarray}
 w_k^{\rm GK} = \frac{1}{\mathcal{N}} \times \frac{1}{N_{\rm rep}} \sum_\ell^{N_{\rm rep}}
 \exp \bigg[
    &  & \hspace{-0.20cm} - \frac{1}{2\Delta \chi^2} \sum_{n,m}^{N_{\rm eig}} R_{n\ell} \left( \sum_{i,j} D_{in} C^{-1}_{ij} D_{jm} \right) R_{m\ell}  
    \label{eq:calc1} \\
    &  & \hspace{-0.20cm} - \frac{1}{\Delta \chi^2} \sum_{n}^{N_{\rm eig}} a_n R_{n\ell} 
 -\frac{1}{2\Delta \chi^2} \, \chi^2[f_{S_0}]
  \bigg]    R_{k\ell}, \nonumber
\end{eqnarray}
where $\chi^2[f_{S_0}]$ is the value of $\chi^2$ computed with the central set $S_0$, and the coefficients $D_{ik}$
and $a_k$ were defined in Eqs.~(\ref{eq:D}) and (\ref{eq:a}). In the limit of infinitely large $N_{\rm rep}$,
the sum over the replicas above can be replaced by an integral
\begin{equation}
 \frac{1}{N_{\rm rep}} \sum_{\ell=1}^{N_{\rm rep}} \xrightarrow{N_{\rm rep} \rightarrow \infty}  \left(2\pi \right)^{-N_{\rm eig}/2}
 \int_{-\infty}^{+\infty} d \vec {\bf R} \exp \left[-\frac{1}{2} \vec {\bf R}^2 \right],
\end{equation}
where the additional exponential stems from the probability distribution for the random numbers
$R_{m\ell}$ being Gaussian. Using this in Eq.~(\ref{eq:calc1}) above, we have
 \begin{eqnarray}
 w_k^{\rm GK} & = & \frac{1}{\mathcal{N}} \times \left(2\pi \right)^{-N_{\rm eig}/2}
 e^{ -\frac{1}{2\Delta \chi^2} \, \chi^2[f_{S_0}] } \int_{-\infty}^{+\infty} d \vec {\bf R}
 e^ { 
 -\frac{1}{2\Delta \chi^2} \vec {\bf R}^T {\bf B} \vec {\bf R} - \frac{1}{\Delta \chi^2} \vec {\bf a}^T \vec {\bf R}
 } R_k \nonumber \\
 & = & - \frac{\Delta\chi^2 }{\mathcal{N}} \times \left(2\pi \right)^{-N_{\rm eig}/2}
 e^{-\frac{1}{2\Delta \chi^2} \, \chi^2[f_{S_0}] }
  \frac{d}{da_k} \int_{-\infty}^{+\infty} d \vec {\bf R}
 e^{ 
 -\frac{1}{2\Delta \chi^2} \vec {\bf R}^T {\bf B} \vec {\bf R} - \frac{1}{\Delta \chi^2} \vec {\bf a}^T \vec {\bf R}
 } \nonumber \\
 & = & - \frac{1}{\mathcal{N}} \times \sqrt{ \frac{(\Delta \chi^2)^{N_{\rm eig}}}{\det {\bf B}} }
 e^{-\frac{1}{2\Delta \chi^2} \, \chi^2[f_{S_0}] + \frac{1}{2\Delta \chi^2} 
 \vec {\bf a}^T {\bf B}^{-1} \vec {\bf a} } ({\bf B}^{-1} \vec {\bf a})_k, \nonumber
 \end{eqnarray}
 where the elements of the matrix ${\bf B}$ were given in Eq.~(\ref{eq:B}). The corresponding
expression for the denominator $\mathcal{N}$ is
\begin{eqnarray}
 \mathcal{N} =  \sqrt{ \frac{(\Delta \chi^2)^{N_{\rm eig}}}{\det {\bf B}} }
 e^{-\frac{1}{2\Delta \chi^2} \, \chi^2[f_{S_0}] + \frac{1}{2\Delta \chi^2} 
 \vec {\bf a}^T {\bf B}^{-1} \vec {\bf a} }.
\end{eqnarray}
Upon taking the ratio, the various prefactors cancel, and the coefficients $w_k^{\rm GK}$ reduce to
\begin{equation}
 w_k^{\rm GK} = -({\bf B}^{-1} \vec {\bf a})_k,
\end{equation}
which coincides with the Eq.~(\ref{eq:wmineq}) specifying the coefficients for the linear Hessian reweighting.
As shown in Ref.~\cite{Sato:2013ika}, the weights $w_k^{\rm chi-squared}$ in Eq.~(\ref{eq:wNNPDF})
emerge from $w_k^{\rm GK}$ by integrating over all possible data sets that give equal $\chi^2$.
Being ``contaminated'' by such additional information readily explains their failure in the present context.

\section{Inclusive jet production at the LHC}

In this section, we apply the reweighting methods to the production of inclusive jets in proton+proton
collisions at the LHC. In comparison to our example discussed earlier, a new source of non-linearity
arises which could potentially decrease the accuracy of the linear reweighting. Namely, the quadratic
PDF dependence of the proton+proton cross sections ($\hat \sigma$ denotes the coefficient functions and jet definitions in general)
\begin{equation}
  \sigma^{\rm pp}[S] = f[S] \otimes \hat \sigma \otimes f[S], 
\end{equation}
may or may not be well approximated by
\begin{eqnarray}
 \sigma^{\rm pp}[S_0 + \frac{1}{2} \sum_k \left(S^+_k - S^-_k \right) w_k] \approx \sigma^{\rm pp}[S_0] + \frac{1}{2} \sum_k \left(\sigma^{\rm pp}[S^+_k] - \sigma^{\rm pp}[S^-_k] \right) w_k,
 \label{eq:ppapprox}
\end{eqnarray}
which is the approximation one effectively makes when using Eq.~(\ref{eq:XS}).
In some other cases, such as
deep-inelastic scattering (linear in PDFs), this would not be an issue whereas for the $W$-asymmetry, the 
non-linearities could be even more intricate. Here, we consider the recent CMS $\sqrt{s}=7 \, {\rm TeV}$ jet
measurements \cite{Chatrchyan:2012bja} ($N_{\rm data}=133$) for which a direct FASTNLO 
interface \cite{Kluge:2006xs,Britzger:2012bs,Wobisch:2011ij} is available to evaluate the cross sections at NLO accuracy.
We use the CTEQ6.6 PDFs \cite{Nadolsky:2008zw} whose error sets quantify the uncertainties within 90\% confidence level the
corresponding tolerance being $\Delta \chi^2_{\rm CTEQ6.6}(90\%) = 100$. 
\begin{figure*}[ht!]
\begin{minipage}[b]{1.00\linewidth}
\centering
\includegraphics[width=0.70\textwidth]{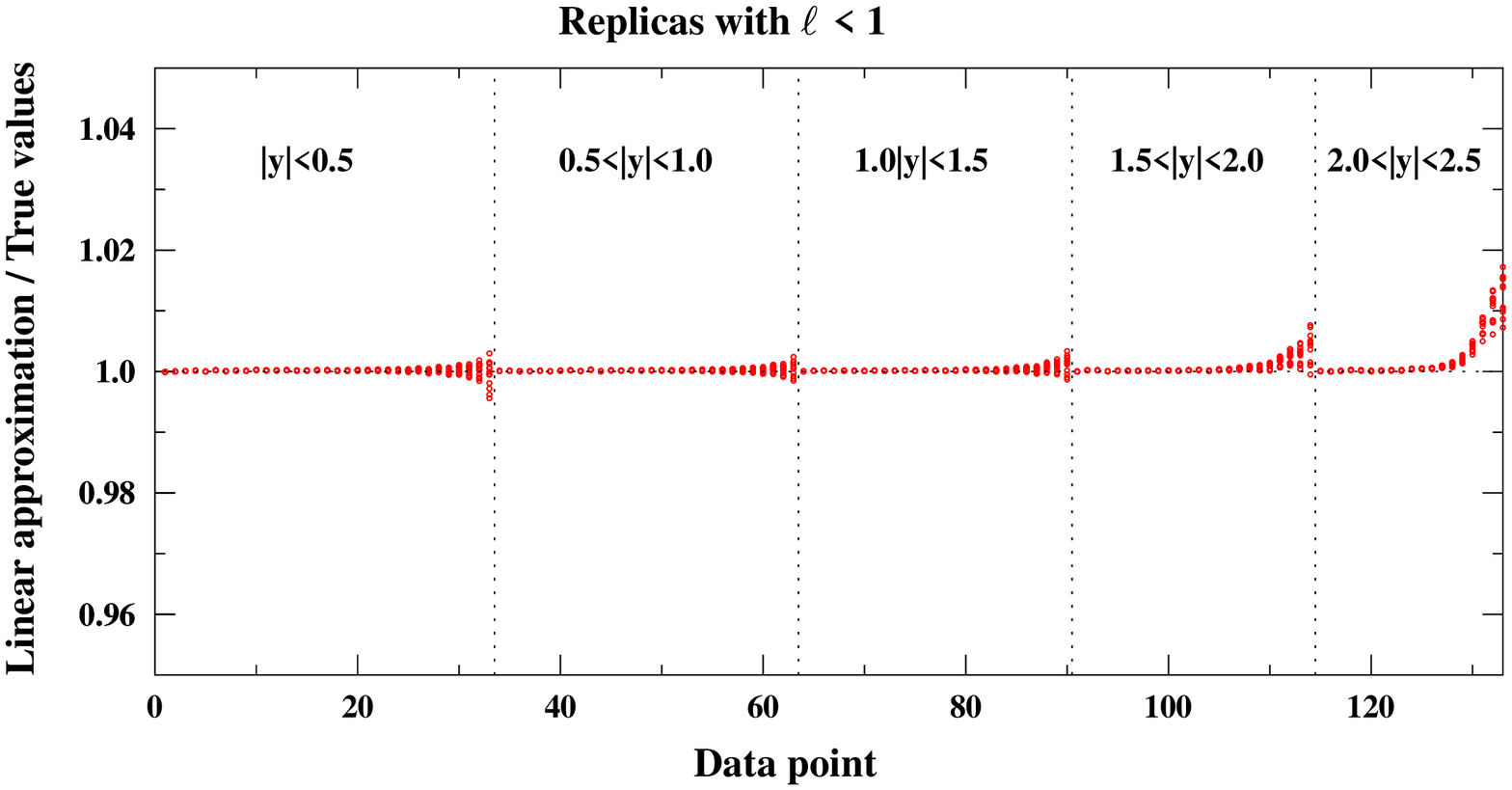}
\includegraphics[width=0.70\textwidth]{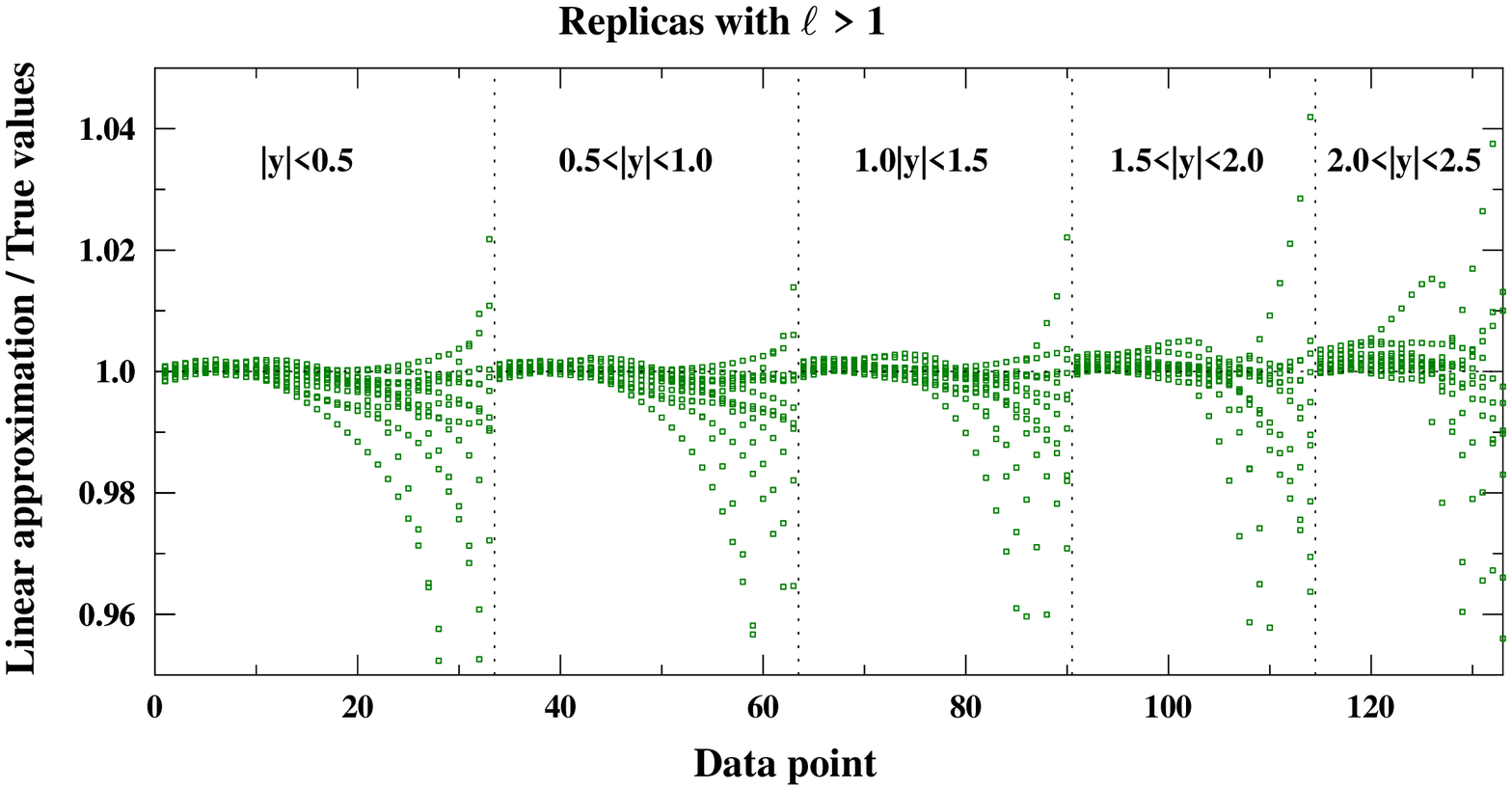}
\end{minipage}
\caption{{\bf Upper panel:} Ratios between the jet cross sections computed by the linear approximation of
Eq.~(\ref{eq:ppapprox}) divided by the ``exact'' value using a parametrization of the corresponding linear 
combination of PDFs inside FASTNLO. The results are plotted for 15 random replicas with the weight-vector
length  $\ell < 1$. The data points have been numbered with increasing transverse momentum. {\bf Lower panel:}
As the upper panel, but with $\ell > 1$.
}
\label{Fig:Linearcheck}
\end{figure*}
The renormalization scale $\mu_r$ and factorization scale $\mu_f$ were fixed to the jet transverse
momentum as $\mu_r=\mu_f=p_T/2$, and the strong coupling was set to $\alpha_s(M_Z)=0.118$ at the Z boson pole.

Let us first discuss the adequacy of the approximation in Eq.~(\ref{eq:ppapprox}). To this end, we have prepared some 
random PDF replicas by Eq.~(\ref{eq:replicas}) separating the cases $\ell = \sum_{i=1}^{N_{\rm eig}}R_{ik}^2 < 1$ and
$\ell = \sum_{i=1}^{N_{\rm eig}}R_{ik}^2 > 1$. We compute the jet cross sections ``exactly'' (by constructing parametrizations
corresponding to these PDF replicas and using them in FASTNLO computations) and, on the other hand, by the linear approximation
of Eq.~(\ref{eq:ppapprox}). Typical results from such an exercise are shown in Figure~\ref{Fig:Linearcheck}. First, if $\ell < 1$, 
the linear approximation proves rather accurate, the deviations being normally much less than couple of percents.
In the latter case, $\ell > 1$, the linear approximation evidently breaks down. We conclude that if the reweighted 
PDFs end up sufficiently close to the original ones the linear approximation of Eq.~(\ref{eq:ppapprox}), and thereby
the linear Hessian reweighting, should be rather accurate. In the case of Bayesian reweighting, replicas with $\ell>1$ often occur, 
and, from the lower panel of Figure~\ref{Fig:Linearcheck}, we can consequently anticipate some differences whether the cross sections are evaluated using the replicas directly inside FASTNLO, or
``on the fly'' by the linear approximation of Eq.~(\ref{eq:ppapprox}). 

\begin{figure*}[ht!]
\begin{minipage}[b]{1.00\linewidth}
\centering
\includegraphics[scale=0.38]{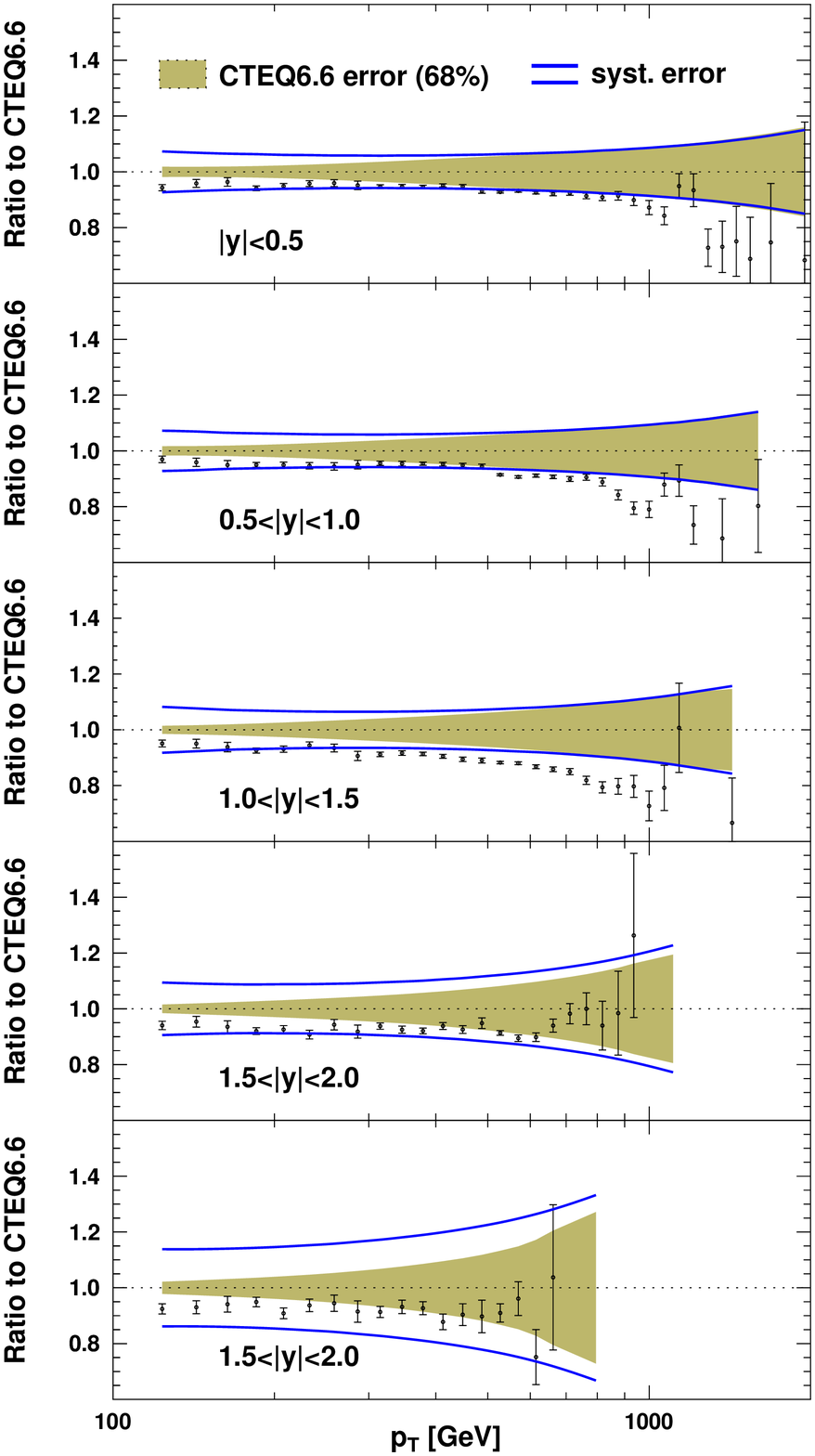} 
\hspace{-0.5cm}
\includegraphics[scale=0.38]{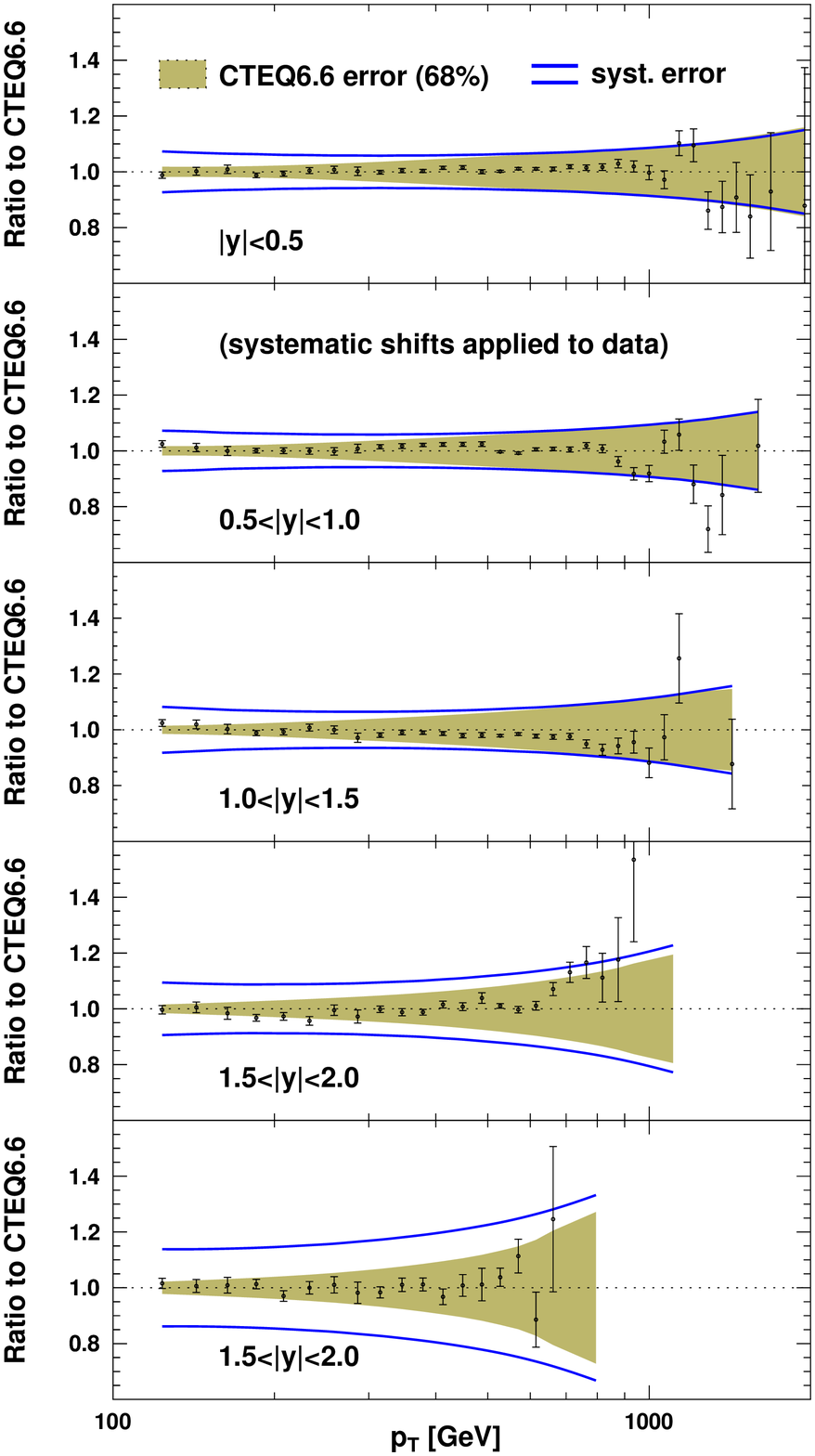}
\end{minipage}
\vspace*{-0.5cm}
\caption{{\bf Left-hand panels:} The CMS inclusive jet cross sections for the five rapidity intervals compared to the
NLO calculation with CTEQ6.6 PDFs and taking $\mu_f=\mu_r=p_T/2$. The error bars in the data points 
show the statistical uncertainty, while the total systematic error is indicated by the blue lines.
The colored bands show the CTEQ6.6 PDF uncertainty. {\bf Right-hand panels:} As the left-hand panels,
but after applying the systematic shifts.}
\label{Fig:CMSdata}
\end{figure*}
We account for the correlated systematic errors by constructing a covariance matrix.
To be specific, we compute the elements of the covariance matrix $C$ by
\begin{equation}
 C_{ij} = \delta_{ij} \left(\sigma_i^{\rm uncorr}\right)^2 + \sum_k \beta_i^k \beta_j^k,
\end{equation}
where $\sigma_i^{\rm uncorr}$ is the uncorrelated error of data point $i$, and $\beta_i^k$ denotes
the absolute shift of this data point corresponding to 1-sigma deviation of the systematic parameter $k$.
In addition to the luminosity, unfolding and jet energy scale uncertainties, we also treat the quoted 
uncertainty in the multiplicative non-perturbative corrections (underlying event, hadronization) as a correlated systematic
error. The uncorrelated errors $\sigma_i^{\rm uncorr}$ include the statistical and 1\% uncorrelated systematic
uncertainty added in quadrature. Calculating the $\chi^2$ using the covariance matrix $C$ is equivalent to
(see e.g. \cite{Gao:2013xoa,Stump:2001gu,Albino:2008fy}) minimizing
\begin{equation}
\chi^2 = \sum_i \left[ \frac{y_i^{\rm theory} - y_i^{\rm data} - \sum_k s_k \beta_i^k}{\sigma_i^{\rm uncorr}} \right]^2 + \sum_k s_k^2,
\label{eq:chi2syst}
\end{equation}
with respect to the systematic parameters $s_k$. This occurs with the parameter values
\begin{equation}
s_k^{\rm min} = \sum_j \left[ \beta_j^k - \sum_{i,\ell,s} \beta_i^k C_{i\ell}^{-1} \beta_\ell^s \beta_j^s \right] \frac{y_j^{\rm theory} - y_j^{\rm data}}{(\sigma_j^{\rm uncorr})^2},
\end{equation}
and $- \sum_k s^{\rm min}_k \beta_i^k$ is the net systematic shift for the data point $y_i^{\rm data}$.
Figure~\ref{Fig:CMSdata} presents a comparison between the CMS data and the NLO predictions, including
the 68\% PDF error bands (obtained downscaling the original 90\% errors by $1/1.645$ \cite{Gauld:2013aja,Dulat:2013kqa}).
The $\chi^2$ value for the central CTEQ6.6 set is $\chi^2/N_{\rm data} \approx 2.1$ 
accounting for the data correlations as discussed above. The calculation appears
to overpredict the experimental cross section by some 5\% which, however, gets still easily hidden under the systematic
shifts as demonstrated in Figure~\ref{Fig:CMSdata}.

\begin{figure*}[htb!]
\begin{minipage}[b]{1.00\linewidth}
\centering
\includegraphics[width=0.70\textwidth]{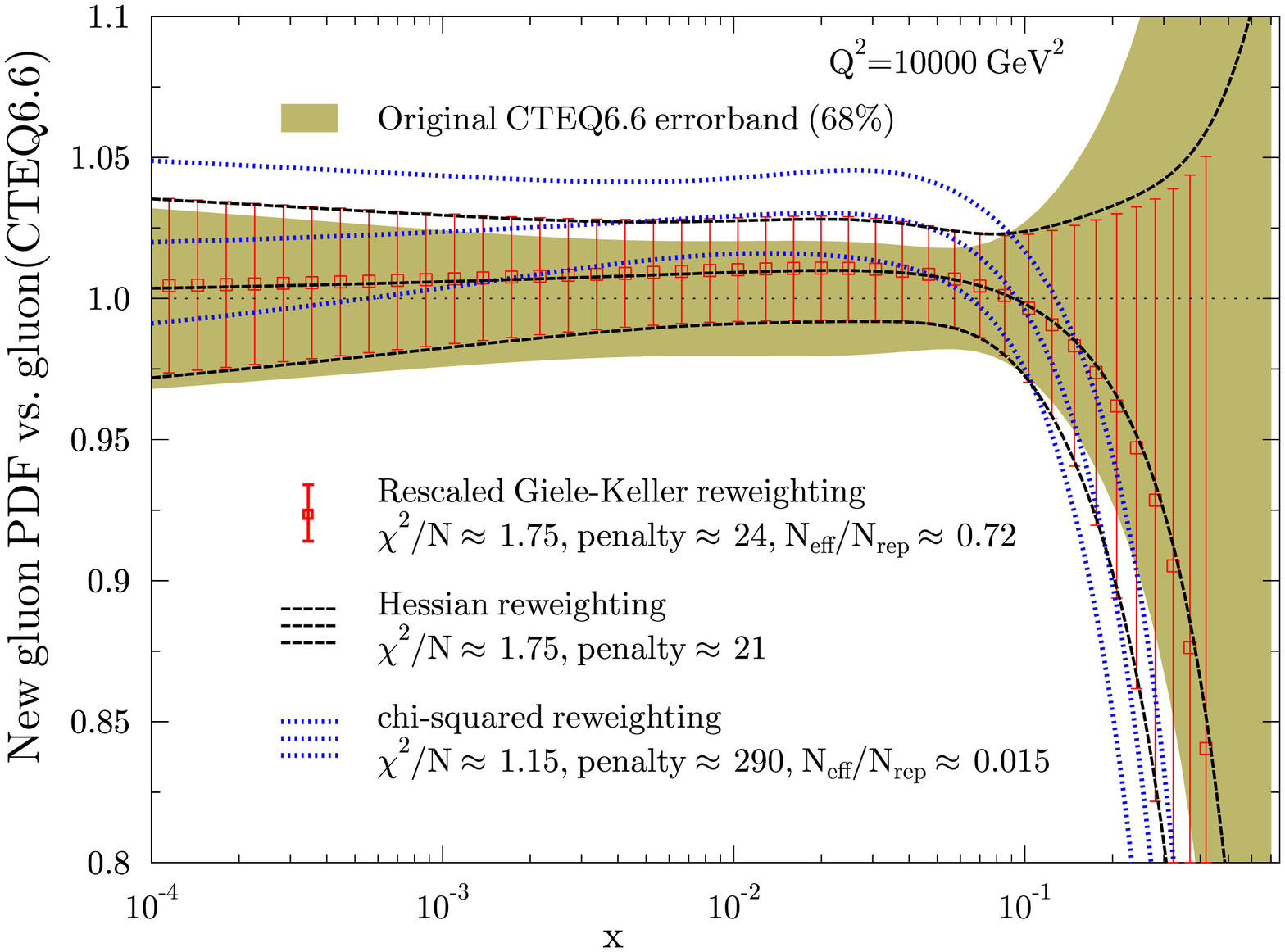}
\includegraphics[width=0.70\textwidth]{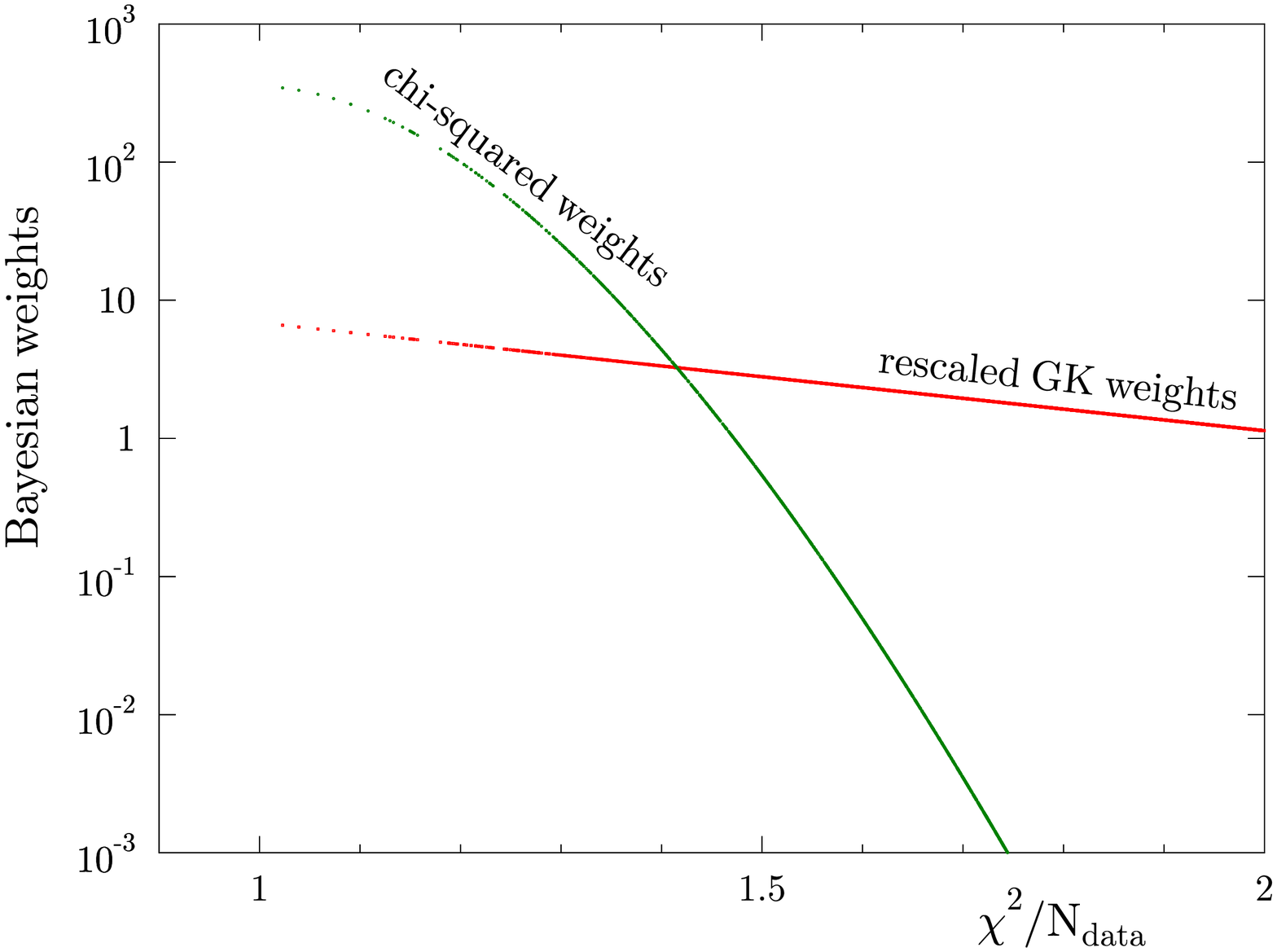}
\end{minipage}
\caption{{\bf Upper panel:} The gluon PDFs at $Q^2=10000 \, {\rm GeV}^2$ after reweighting using the
correlated errors. The red points with error bars correspond to the results using the Bayesian reweighting with rescaled GK weights,
and the blue dotted lines mark the corresponding result with chi-squared weights. The result from Hessian reweighting
is indicated by the black dashed lines and the colored band is the original CTEQ6.6 uncertainty. All results are 
normalized to the central set of CTEQ6.6. {\bf Lower panel:} The distribution of Bayesian weights.
}
\label{Fig:NewGluons}
\end{figure*}

It is a straightforward task to apply the reweighting methods on these jet data. For the needs of
the Bayesian techniques we have generated $10^4$ PDF replicas using Eq.~(\ref{eq:replicas}).
As the data uncertainties correspond to one standard deviation, we have rescaled the random numbers in Eq.~(\ref{eq:replicas}) 
by $R_{ik} \rightarrow R_{ik}/1.645$ which brings the PDF replicas to the 68\% level as well. When computing the GK weights
we divide the resulting values of $\chi^2$ by $\Delta \chi^2_{\rm CTEQ6.6}(68\%)=100/1.645^2\approx 37$, to
appropriately modify the underlying likelihood. The chi-squared weights are always computed without rescaling 
the $\chi^2$ values. We stress that the rescaling of CTEQ6.6 replicas from 90\% to 68\% confidence level affects only
the outcome of chi-squared reweighting as, by construction, the central result of GK reweighting would remain the same
by simply using the 90\% replicas and rescaling the $\chi^2$ values by $\Delta \chi^2_{\rm CTEQ6.6}(90\%)$.
The Hessian reweighting has been performed directly with the original 90\% error sets rescaling the uncertainties 
by $1/1.645$ only at the end. The resulting modifications in the gluon PDFs
are presented in the upper panel of Figure~\ref{Fig:NewGluons} (as the jet production is predominantly
sensitive to the gluons, we find it reasonable to present the results here only in terms of gluon PDFs). 
From this plot we find that, as expected, the Hessian technique and the Bayesian method with 
rescaled GK weights are in good agreement also here. The small mismatch at large $x$ originates mainly 
from Eq.~(\ref{eq:ppapprox}) not being exact. As the reweighting penalty $P \approx 21$ is 
less than $\Delta \chi^2_{\rm CTEQ6.6}(68\%)$, we conclude that these data could have 
been added to the CTEQ6.6 fit without causing a significant disagreement with the original data. In this sense these
data are compatible with the CTEQ6.6 PDFs despite the largish $\chi^2/N_{\rm data} \approx 1.75$ which could hint,
however, that some tension between the Tevatron Run-1 jet data \cite{Affolder:2001fa,Abbott:2000kp} (used in CTEQ6.6
fit to constrain large-$x$ gluons) and these new LHC measurements exist.\footnote{Indeed, the latest CTEQ fit \cite{Gao:2013xoa} 
has abandoned the Tevatron Run-1 jet data as they do not completely agree with the Run-2 data. See Ref.~\cite{Martin:2009iq}
for a review and  Ref.~\cite{Pumplin:2009nk} for more discussions.}
\begin{figure*}[ht!]
\begin{minipage}[b]{1.00\linewidth}
\centering
\includegraphics[scale=0.38]{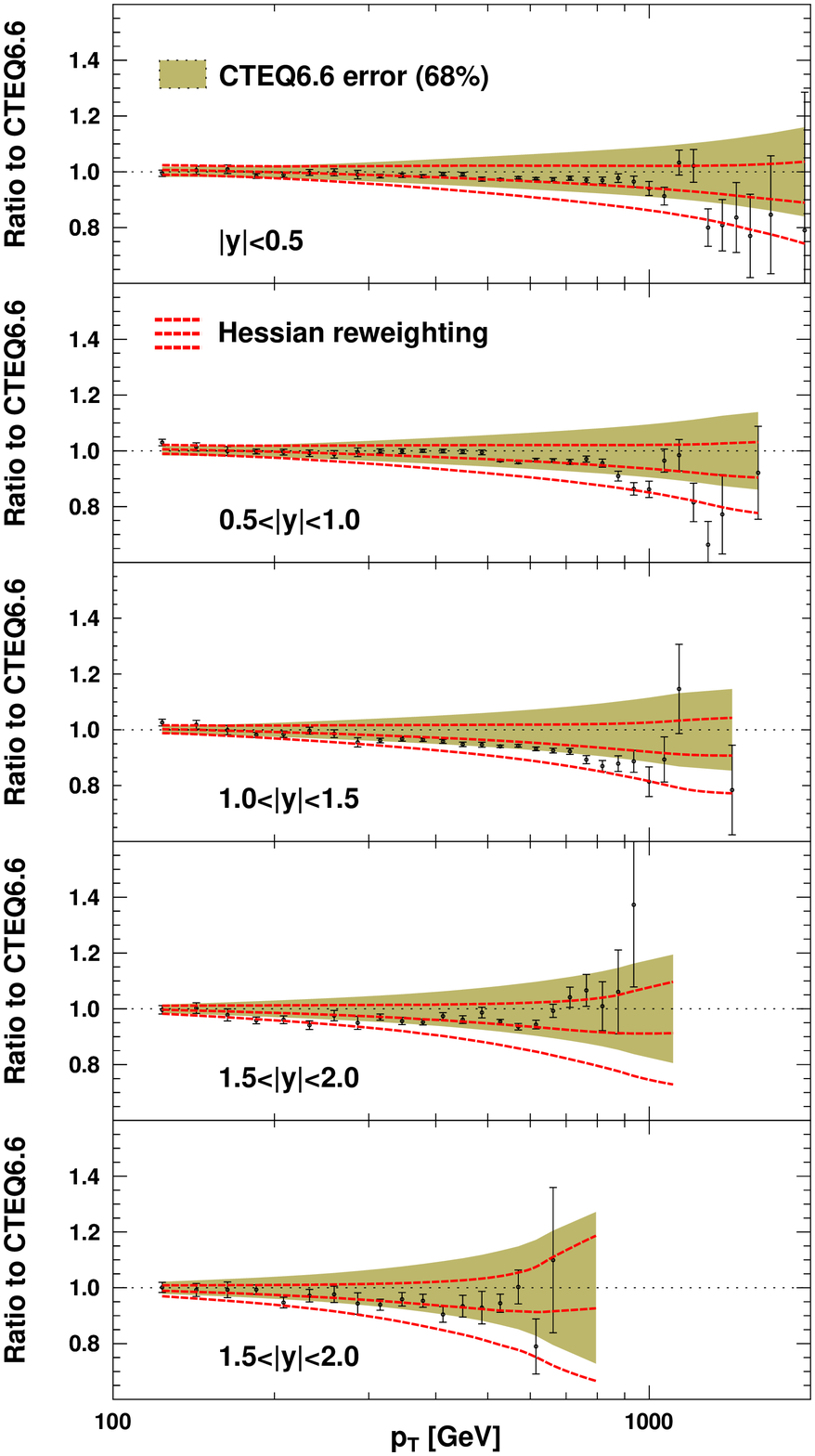} 
\hspace{-0.5cm}
\includegraphics[scale=0.38]{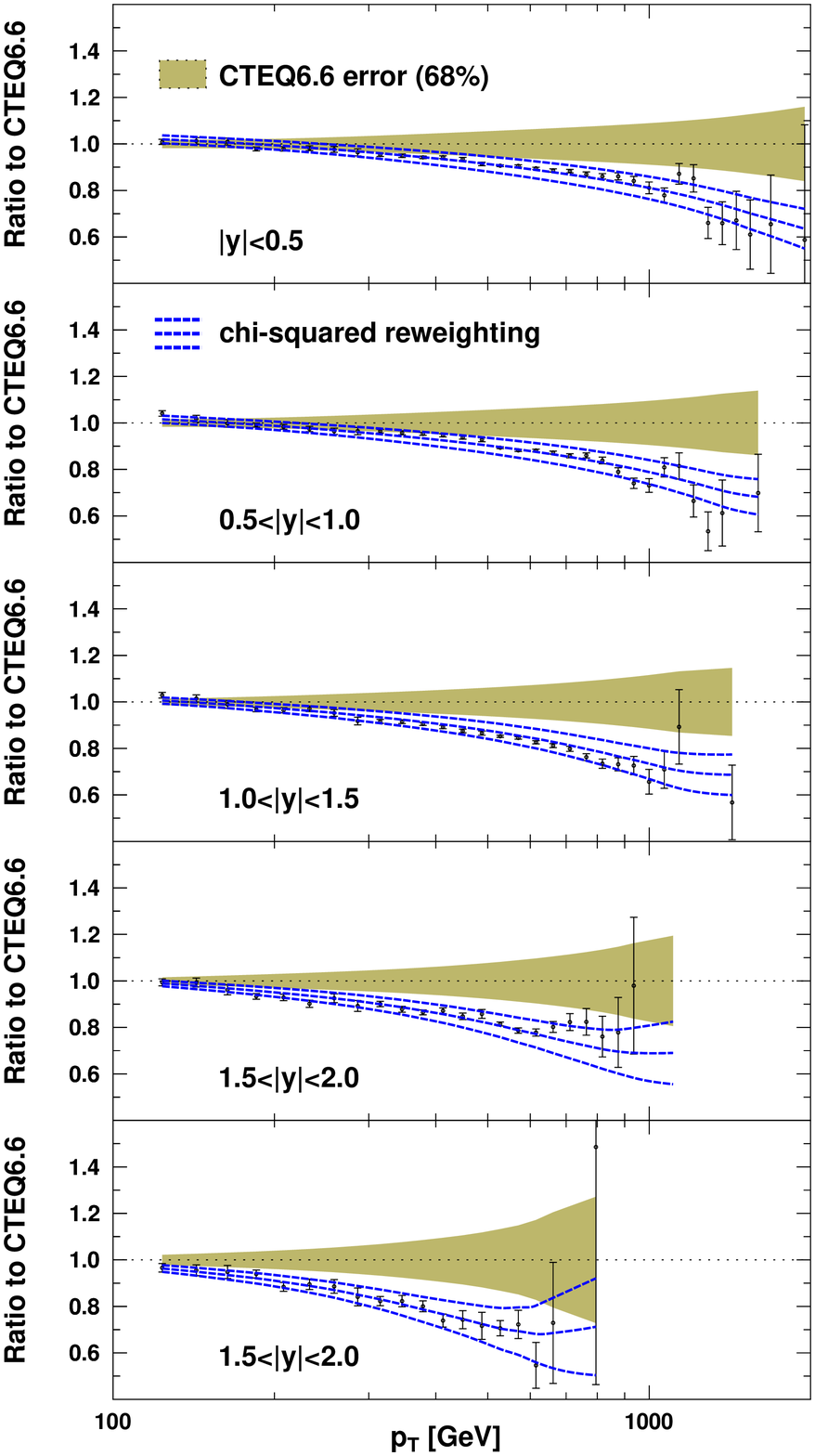}
\end{minipage}
\caption{{\bf Left-hand panels:} The CMS inclusive jet cross sections compared to the predictions
after the Hessian reweighting and applying the systematic shifts. All values have been normalized to
the central prediction of CTEQ6.6 and only the statistical data errors are shown. {\bf Right-hand panels:}
As the left-hand panels, but with the Bayesian reweighting with chi-squared weights.}
\label{Fig:CMSdata2}
\end{figure*}
The line of red points in Figure~\ref{Fig:NewGluons} shows the spectrum of rescaled GK weights.
For dividing the values of $\chi^2$ by the tolerance $\Delta \chi^2_{\rm CTEQ6.6}(68\%) \approx 37$, this distribution
is rather flat and the final result is affected by replicas with a wide range of $\chi^2$ explaining also the
rather large number of effective replicas, $N_{\rm eff}/N_{\rm rep}~ \approx 0.72$.
The chi-squared weights give rise to effects which are qualitatively alike but quantitatively much larger.
The agreement with the new jet data is admittedly better, $\chi^2/N_{\rm data} \approx 1.15$ 
(the green points in Figure~\ref{Fig:NewGluons} indeed demonstrate that the chi-squared weights assign
clearly larger weights to replicas with $\chi^2/N_{\rm data}<1.4$ than the rescaled GK weights), but at the cost of increasing the original
$\chi^2$ by $P \approx 290$ which is way beyond the CTEQ6.6 tolerance (both 68\% and 90\%).
The surviving number of replicas $N_{\rm eff}$ is also very low. That is, one
could conclude that the jet data considered here were not in agreement with the CTEQ6.6 PDFs. However, as we have
demonstrated in the previous sections, the result is misleading since the chi-squared weights do not appropriately
account for the other data that were originally included in the global CTEQ6.6 fit. 
From Eq.~(\ref{eq:opimumchi2}) we can estimate the value of $\Delta \chi^2$ that would have brought the results
of chi-squared reweighting close to the correct ones: using the original central value
$\chi^2/N_{\rm data} \approx 2.1$ and $N_{\rm data}=133$ we obtain $\Delta \chi^2 \approx 1.9$. This is much less
than $\Delta \chi^2_{\rm CTEQ6.6}(68\%)=37$ and concretely explains why the chi-squared weights lead to a result
which is so far from the correct one.

The new predictions for the cross-section are shown in Figure~\ref{Fig:CMSdata2}. The systematic shifts have 
been applied to the data and they depend significantly on which method of reweighting was used. The Hessian
reweighting has caused a mild downward shift on the cross section which was to be expected given that the
original central predictions somewhat overshoot the data (see Figure~\ref{Fig:CMSdata}).
In the case of Bayesian reweighting with chi-squared weights the induced changes with respect to the original
CTEQ6.6 predictions are way too large.

\subsection*{Reweighting of MSTW2008}

In the previous example, we discussed PDF reweighting in the case of a fixed global tolerance $\Delta \chi^2$ assuming the
validity of the quadratic approximation in Eq.~(\ref{eq:newchi2}). It is, however, well known that such
quadratic profile is never perfect, and for the fit parameters that are not very well constrained
severe deviations from the ideal quadratic behaviour may occur (see e.g. Figs.~5-6 in \cite{Martin:2009iq}). To
account for such imperfections and improve the linear approximations in estimating the observables in the
space of eigenvectors, the non-linear extension of the Hessian reweighting should provide some improvement.
As an example, we consider the MSTW2008 PDFs \cite{Martin:2009iq} using the same CMS jet data as earlier.
In fact, the effect of incorporating the $\sqrt{s}=7 \, {\rm TeV}$ CMS and ATLAS \cite{Aad:2011fc} inclusive jet measurements into the MSTW2008 framework was
studied recently via a direct re-fit \cite{Watt:2013oha}, but due to the large systematic uncertainties of the
ATLAS data, the fit was mainly driven by the same CMS measurements we have discussed here. Thus, in what follows,
we will contrast our results to what was obtained in that analysis.

\begin{figure*}[ht!]
\begin{minipage}[b]{1.00\linewidth}
\centering
\includegraphics[width=0.70\textwidth]{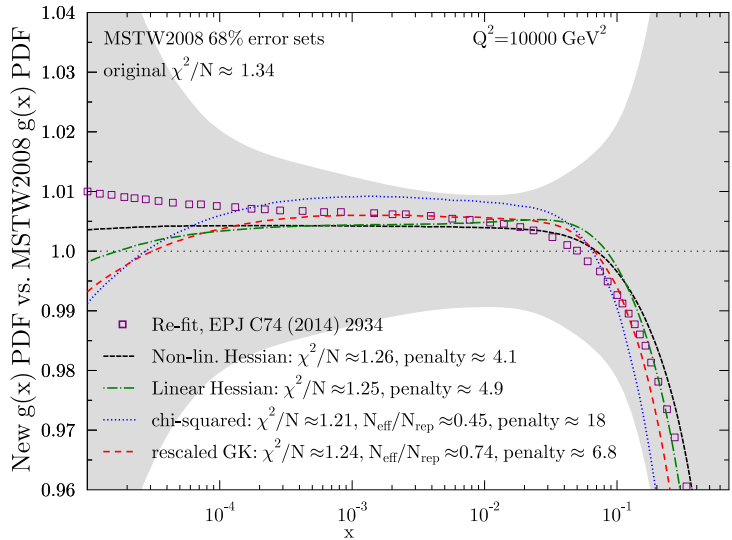}
\end{minipage}
\caption{
The reweighted gluon distributions at $Q^2=10000 \, {\rm GeV}^2$  normalized by 
the central MSTW2008 gluons. The solid black line is the result from the non-linear
Hessian reweighting, and the green dashed-dotted line corresponds to the linear Hessian reweighting. Blue dotted 
line is the prediction of Bayesian reweighting with chi-squared weights, and red dashed line results from using
the rescaled GK weights. The shaded band is the original MSTW2008 uncertainty and the purple squares mark the
result obtained in Ref.~\cite{Watt:2013oha}.
}
\label{Fig:MSTWgluons1}
\end{figure*}

The MSTW2008 package provides two separate sets of error PDFs, the 68\% and 90\% confidence-level sets,
of which we will use the former one. As the numerical values for the determined tolerances $t^\pm_k$ and the corresponding 
increases $T^\pm_k$ of the global $\chi^2$ are given, we can account also for the non-quadratic
behaviour of the original $\chi^2$ function in the space of eigenvectors. Specifically, in addition to using
Eq.~(\ref{eq:nonliny}) in evaluating the observables, we make the following substitution in Eq.~(\ref{eq:newchi2}):
\begin{eqnarray}
\sum_k^{N_{\rm eig}} z_k^2 & \longrightarrow & \sum_k^{N_{\rm eig}} a_k z_k^2 + b_k z_k^3 \, , \nonumber \\
a_k & = & \frac{t_k^-(T_k^+/t_k^+)^2  - t_k^+(T_k^-/t_k^-)^2}{t_k^+ +t_k^-} \, , \\
b_k & = & \frac{(T_k^+/t_k^+)^2 - (T_k^-/t_k^-)^2}{t_k^+ +t_k^-} \, . \nonumber
\end{eqnarray}
We set $\alpha_s(M_Z^2)=0.1202$, and use $\mu_r=\mu_f=p_T$. As it is not possible
(with the available information) to consistently implement the dynamic tolerance 
after the reweighting, we discuss here only the resulting central values.

\begin{figure*}[ht!]
\begin{minipage}[b]{1.00\linewidth}
\centering
\includegraphics[scale=0.38]{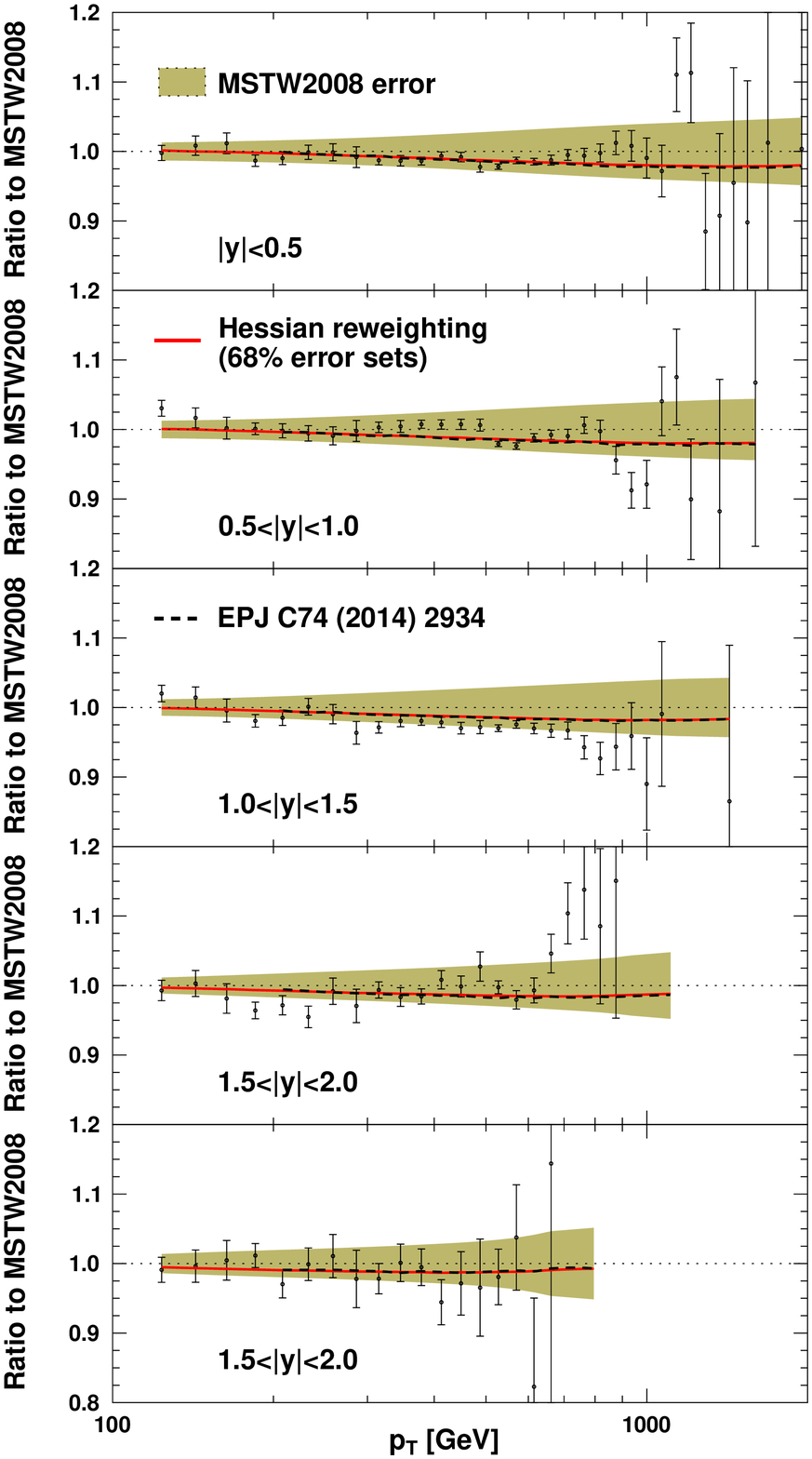} 
\hspace{-0.5cm}
\includegraphics[scale=0.38]{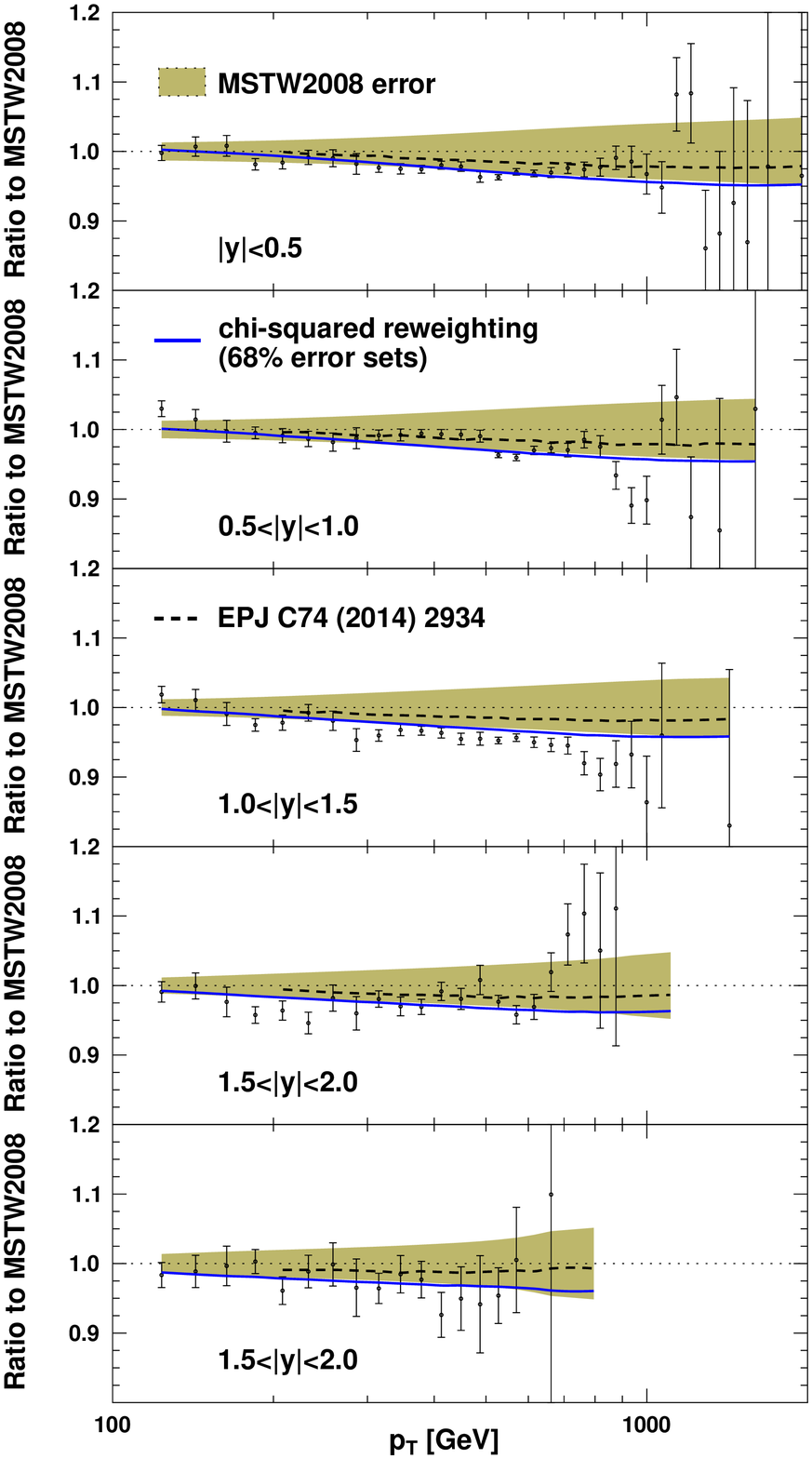}
\end{minipage}
\vspace*{-0.5cm}
\caption{
{\bf Left-hand panels:} The jet cross sections after reweighting MSTW2008 PDFs.
The results of the non-linear Hessian reweighting are shown by the red lines and the results of Ref.~\cite{Watt:2013oha}
are the black dashed lines. All values have been normalized to the central predictions of MSTW2008.
Systematic shifts have been applied to the data and only the statistical data errors are shown. 
{\bf Right-hand panels:} As the left-hand panels, but with the Bayesian reweighting with chi-squared weights.}
\label{Fig:CMSdata3}
\end{figure*}

The gluon PDF after the non-linear Hessian reweighting is shown in Figure~\ref{Fig:MSTWgluons1}. For comparison,
the figure includes also the outcome from the linear Hessian reweighting, Bayesian reweighting with the chi-squared
weights, Bayesian reweighting with GK weights rescaled by the average tolerance, $\Delta \chi^2_{\rm MSTW2008}(68\%) \approx 10$,
and the result obtained by a re-fit in Ref.~\cite{Watt:2013oha}.
Apart from the Bayesian predictions with chi-squared weights the results of different prescriptions are more or
less consistent with each other within $10^{-4}<x<10^{-1}$,
and also close to the ``exact'' result.\footnote{
Note that in Ref.~\cite{Watt:2013oha} the open fit parameters were not restricted to those used in generating
the 20 eigenvecor sets. Thus, it is natural to expect deviations in comparison to the results of reweighting.}
The differences between the linear and non-linear prescriptions are clearly more important
at $x<10^{-4}$ and $x>10^{-1}$, where the original PDF uncertainties become large.
The chi-squared weights predict again modifications for the gluon PDFs  which are similar 
to the correct ones but somewhat too pronounced (the estimated penalty stands out from what is obtained with the other methods). 
The reason can be understood by estimating from Eq.~(\ref{eq:opimumchi2}) which value of $\Delta \chi^2$ would have given 
consistent results: now $\chi^2/N_{\rm data} \approx 1.34$ which translates
to $\Delta \chi^2 \approx 3.9$. While close, this is still less than $\Delta \chi^2_{\rm MSTW2008}(68\%) \approx 10$ 
and explains why the result is different from the direct re-fit. 

The predicted jet cross sections after the reweighting are shown in Figure~\ref{Fig:CMSdata3},
normalized to the central MSTW2008 values. The results of Hessian reweighting clearly stay within the original uncertainties
and appear to agree with the results of Ref.~\cite{Watt:2013oha}. The predictions of the Bayesian
reweighting with chi-squared weights are, however, at the lower limit of the original error bands
and disgaree with the true re-fit.

\section{Conclusion}

We have discussed how to test the consistency and estimate the effects that a new set of experimental 
measurements have within an existing set of Hessian PDFs with non-zero
tolerance $\Delta \chi^2$. To this end, the Hessian reweighting, was introduced as an alternative technique to the
prevailing Bayesian methods. While the Hessian reweighting is straightforwardly derived considering
a new set of data in a global $\chi^2$ fit, the Bayesian methods are outwardly distinct, based on statistical
inference. We compared the different approaches to a direct re-fit through a simple example verifying
the adequacy of the Hessian method. In the case of the Bayesian procedure an agreement with a new fit
was also established, but only after including the $\Delta \chi^2$ criterion properly into the Bayesian
likelihood function which --- as we mathematically justified --- must be a pure exponential as originally
proposed by Giele and Keller. 
The conditions under which another commonly used (but in this case inadequate) likelihood function gives consistent
results was also discussed.
The inclusive jet production at the LHC was considered as an additional example.
At first, our findings may appear to be in contrast to the works of NNPDF collaboration in which a different
functional form for the likelihood is derived. However, as the NNPDF methodology for fitting PDFs is far more
involved than the simple $\chi^2$ minimization considered here, it is possible that a different functional form 
applies. 
For example, the NNPDF fits involve extremely flexible fit functions and to avoid fitting random fluctuations
the data are divided into ``training`` and ''control'' sets. While the actual $\chi^2$ minimization is done only
for the former subset of data, the latter one is used to decide when the minimization process should be stopped. 
The outcome of a direct $\chi^2$ minimization with no stopping criteria would generally be different
if the same functional form would be used. However, for the moment it is still unclear what exactly causes the need for 
using a different Bayesian likelihood.

The two types of methods discussed here have their pros and cons: While the Hessian procedure requires evaluating
the observables only with the central and error sets (typically around 50 sets in total), in the Bayesian
method one needs to deal with a much larger ensemble of PDFs (around $10^3$ to find well-converging results).
On the other hand, the Hessian reweighting is procedurally a bit more involved requiring e.g. numeric linear
algebra, while the Bayesian technique is simpler. The reliability of the both methods depends basically on the
accuracy of the quadratic approximation around the minimum $\chi^2$ made in the original PDF fit and on the
adequacy of the linear approximations that one makes. In the case of Hessian reweighting one can improve on
these approximations by invoking non-linear corrections as we explained. However, when the PDF fits are
updated details like the form of the fit function, data sets included, or value of $\alpha_s(M_Z^2)$ are
often altered. Such modifications cannot be easily accounted for by the reweighting procedures and in this sense the
reweighting is always an approximation to a real fit. In any case, it is definitely useful in checking whether
the new data appear consistent and which components of the PDFs are bound to undergo a change and how.

\section*{Acknowledgments}
We thank J.~Rojo and S.~Forte for useful discussions on the PDF reweighting and N.~Armesto and C.~Salgado for their comments regarding
the manuscript. We also want to acknowledge the Referee of our paper whose suggestions led to significant 
improvements especially in Section 5 concerning the discussion on the conditions that make the chi-squared reweighting
to closely mimic a direct refit.
During this work H.P. was supported by the Academy of Finland, Project No. 133005. P.Z. was supported by 
European Research Council grant HotLHC ERC-2011-StG-279579; by Ministerio de Ciencia e Innovaci\'on of Spain
under the Consolider-Ingenio 2010 Programme CPAN (CSD2007-00042);
by Xunta de Galicia; and by FEDER.


\begin{thebibliography}{99}

\bibitem{Collins:1989gx}
  J.~C.~Collins, D.~E.~Soper and G.~F.~Sterman,
  Adv.\ Ser.\ Direct.\ High Energy Phys.\  {\bf 5} (1988) 1
  [hep-ph/0409313].

\bibitem{Collins:1985ue}
  J.~C.~Collins, D.~E.~Soper and G.~F.~Sterman,
  Nucl.\ Phys.\ B {\bf 261} (1985) 104.
  
\bibitem{Forte:2013wc}
  S.~Forte and G.~Watt,
  Ann.\ Rev.\ Nucl.\ Part.\ Sci.\  {\bf 63} (2013) 291
  [arXiv:1301.6754 [hep-ph]].
  
\bibitem{Lai:2010vv}
  H.~-L.~Lai, M.~Guzzi, J.~Huston, Z.~Li, P.~M.~Nadolsky, J.~Pumplin, C.~-P.~Yuan and ,
  Phys.\ Rev.\ D {\bf 82} (2010) 074024
  [arXiv:1007.2241 [hep-ph]].

\bibitem{Martin:2009iq}
  A.~D.~Martin, W.~J.~Stirling, R.~S.~Thorne and G.~Watt,
  Eur.\ Phys.\ J.\ C {\bf 63} (2009) 189
  [arXiv:0901.0002 [hep-ph]].

\bibitem{Ball:2012cx}
  R.~D.~Ball, V.~Bertone, S.~Carrazza, C.~S.~Deans, L.~Del Debbio, S.~Forte, A.~Guffanti and N.~P.~Hartland {\it et al.},
  Nucl.\ Phys.\ B {\bf 867} (2013) 244
  [arXiv:1207.1303 [hep-ph]].

\bibitem{Giele:1998gw}
  W.~T.~Giele and S.~Keller,
  Phys.\ Rev.\ D {\bf 58} (1998) 094023
  [hep-ph/9803393].

\bibitem{Ball:2010gb}
  R.~D.~Ball {\it et al.}  [NNPDF Collaboration],
  Nucl.\ Phys.\ B {\bf 849} (2011) 112
   [Erratum-ibid.\ B {\bf 854} (2012) 926]
   [Erratum-ibid.\ B {\bf 855} (2012) 927]
  [arXiv:1012.0836 [hep-ph]].

\bibitem{Ball:2011gg}
  R.~D.~Ball, V.~Bertone, F.~Cerutti, L.~Del Debbio, S.~Forte, A.~Guffanti, N.~P.~Hartland and J.~I.~Latorre {\it et al.},
  Nucl.\ Phys.\ B {\bf 855} (2012) 608
  [arXiv:1108.1758 [hep-ph]].

\bibitem{Chatrchyan:2013mza}
  S.~Chatrchyan {\it et al.}  [CMS Collaboration],
  Phys.\ Rev.\ D {\bf 90} (2014) 032004
  [arXiv:1312.6283 [hep-ex]].
  
\bibitem{Gauld:2013aja}
  R.~Gauld,
  JHEP {\bf 1402} (2014) 126
  [arXiv:1311.1810 [hep-ph]].

\bibitem{Czakon:2013tha}
  M.~Czakon, M.~L.~Mangano, A.~Mitov and J.~Rojo,
  JHEP {\bf 1307} (2013) 167
  [arXiv:1303.7215 [hep-ph]].

\bibitem{Carminati:2012mm}
  L.~Carminati, G.~Costa, D.~D'Enterria, I.~Koletsou, G.~Marchiori, J.~Rojo, M.~Stockton and F.~Tartarelli,
  EPL {\bf 101} (2013) 61002
   [Europhys.\ Lett.\  {\bf 101} (2013) 61002]
  [arXiv:1212.5511].
  
\bibitem{Beneke:2012wb}
  M.~Beneke, P.~Falgari, S.~Klein, J.~Piclum, C.~Schwinn, M.~Ubiali and F.~Yan,
  JHEP {\bf 1207} (2012) 194
  [arXiv:1206.2454 [hep-ph]].

\bibitem{d'Enterria:2012yj}
  D.~d'Enterria and J.~Rojo,
  Nucl.\ Phys.\ B {\bf 860} (2012) 311
  [arXiv:1202.1762 [hep-ph]].

\bibitem{Ball:2013hta}
  R.~D.~Ball {\it et al.}  [NNPDF Collaboration],
  Nucl.\ Phys.\ B {\bf 877} (2013) 2,  290
  [arXiv:1308.0598 [hep-ph]].
  
\bibitem{Pumplin:2001ct}
  J.~Pumplin, D.~Stump, R.~Brock, D.~Casey, J.~Huston, J.~Kalk, H.~L.~Lai, W.~K.~Tung,
  Phys.\ Rev.\ D {\bf 65} (2001) 014013
  [hep-ph/0101032].

\bibitem{Watt:2012tq}
  G.~Watt, R.~S.~Thorne,
  JHEP {\bf 1208} (2012) 052
  [arXiv:1205.4024 [hep-ph]].

\bibitem{Sato:2013wea}
  N.~Sato,
  arXiv:1309.7995 [hep-ph].

\bibitem{Armesto:2013kqa}
  N.~Armesto, J.~Rojo, C.~A.~Salgado and P.~Zurita,
  JHEP {\bf 1311} (2013) 015
  [arXiv:1309.5371 [hep-ph]].

\bibitem{Watt:2013oha}
  B.~J.~A.~Watt, P.~Motylinski and R.~S.~Thorne,
  Eur.\ Phys.\ J.\ C {\bf 74} (2014) 2934
  [arXiv:1311.5703 [hep-ph]].

\bibitem{Paukkunen:2013grz}
  H.~Paukkunen and C.~A.~Salgado,
  Phys.\ Rev.\ Lett.\  {\bf 110} (2013) 212301
  [arXiv:1302.2001 [hep-ph]].

\bibitem{Sato:2013ika}
  N.~Sato, J.~F.~Owens and H.~Prosper,
  Phys.\ Rev.\ D {\bf 89} (2014) 114020
  [arXiv:1310.1089 [hep-ph]].
  
\bibitem{CooperSarkar:2011aa}
  A.~M.~Cooper-Sarkar [ZEUS and H1 Collaborations],
  PoS EPS {\bf -HEP2011} (2011) 320
  [arXiv:1112.2107 [hep-ph]].

\bibitem{Alekhin:2012ig}
  S.~Alekhin, J.~Blumlein and S.~Moch,
  Phys.\ Rev.\ D {\bf 86} (2012) 054009
  [arXiv:1202.2281 [hep-ph]].

\bibitem{Owens:2012bv}
  J.~F.~Owens, A.~Accardi and W.~Melnitchouk,
  Phys.\ Rev.\ D {\bf 87} (2013) 094012
  [arXiv:1212.1702 [hep-ph]].

\bibitem{Pumplin:2002vw}
  J.~Pumplin, D.~R.~Stump, J.~Huston, H.~L.~Lai, P.~M.~Nadolsky and W.~K.~Tung,
  JHEP {\bf 0207} (2002) 012
  [hep-ph/0201195].

\bibitem{Pumplin:2009bb}
  J.~Pumplin,
  Phys.\ Rev.\ D {\bf 82} (2010) 114020
  [arXiv:0909.5176 [hep-ph]].

\bibitem{Glazov:2010bw}
  A.~Glazov, S.~Moch and V.~Radescu,
  Phys.\ Lett.\ B {\bf 695} (2011) 238
  [arXiv:1009.6170 [hep-ph]].

\bibitem{Chatrchyan:2012bja}
  S.~Chatrchyan {\it et al.}  [CMS Collaboration],
  Phys.\ Rev.\ D {\bf 87} (2013) 112002
  [arXiv:1212.6660 [hep-ex]].

\bibitem{Kluge:2006xs}
  T.~Kluge, K.~Rabbertz and M.~Wobisch,
  hep-ph/0609285.

\bibitem{Britzger:2012bs}
  D.~Britzger {\it et al.}  [fastNLO Collaboration],
  arXiv:1208.3641 [hep-ph].

\bibitem{Wobisch:2011ij}
  M.~Wobisch {\it et al.}  [fastNLO Collaboration],
  arXiv:1109.1310 [hep-ph].

\bibitem{Nadolsky:2008zw}
  P.~M.~Nadolsky, H.~-L.~Lai, Q.~-H.~Cao, J.~Huston, J.~Pumplin, D.~Stump, W.~-K.~Tung and C.~-P.~Yuan,
  Phys.\ Rev.\ D {\bf 78} (2008) 013004
  [arXiv:0802.0007 [hep-ph]].

\bibitem{Gao:2013xoa}
  J.~Gao, M.~Guzzi, J.~Huston, H.~-L.~Lai, Z.~Li, P.~Nadolsky, J.~Pumplin and D.~Stump {\it et al.},
  arXiv:1302.6246 [hep-ph].

\bibitem{Stump:2001gu}
  D.~Stump, J.~Pumplin, R.~Brock, D.~Casey, J.~Huston, J.~Kalk, H.~L.~Lai and W.~K.~Tung,
  Phys.\ Rev.\ D {\bf 65} (2001) 014012
  [hep-ph/0101051].

\bibitem{Albino:2008fy}
  S.~Albino, B.~A.~Kniehl and G.~Kramer,
  Nucl.\ Phys.\ B {\bf 803} (2008) 42
  [arXiv:0803.2768 [hep-ph]].

\bibitem{Dulat:2013kqa}
  S.~Dulat, T.~J.~Hou, J.~Gao, J.~Huston, P.~Nadolsky, J.~Pumplin, C.~Schmidt and D.~Stump {\it et al.},
  Phys.\ Rev.\ D {\bf 89} (2014) 113002
  [arXiv:1310.7601 [hep-ph]].
  
\bibitem{Affolder:2001fa}
  T.~Affolder {\it et al.}  [CDF Collaboration],
  Phys.\ Rev.\ D {\bf 64} (2001) 032001
   [Erratum-ibid.\ D {\bf 65} (2002) 039903]
  [hep-ph/0102074].
  
\bibitem{Abbott:2000kp}
  B.~Abbott {\it et al.}  [D0 Collaboration],
  Phys.\ Rev.\ D {\bf 64} (2001) 032003
  [hep-ex/0012046].

\bibitem{Pumplin:2009nk}
  J.~Pumplin, J.~Huston, H.~L.~Lai, P.~M.~Nadolsky, W.~-K.~Tung and C.~-P.~Yuan,
  Phys.\ Rev.\ D {\bf 80} (2009) 014019
  [arXiv:0904.2424 [hep-ph]].

\bibitem{Aad:2011fc}
  G.~Aad {\it et al.}  [ATLAS Collaboration],
  Phys.\ Rev.\ D {\bf 86} (2012) 014022
  [arXiv:1112.6297 [hep-ex]].

 
\end{thebibliography}
\end{document}